\documentclass[a4paper,fleqn]{cas-dc}

\usepackage[numbers,sort&compress]{natbib}
\usepackage{amsmath,amssymb}
\usepackage{siunitx}
\DeclareSIUnit\rydberg{Ry}
\DeclareSIUnit\bohrmagneton{\ensuremath{\mu_\mathrm{B}}}
\begin{document}
\let\WriteBookmarks\relax
\def\floatpagepagefraction{1}
\def\textpagefraction{.001}

\shorttitle{Pyrazinic N-Doped Porous Graphene Nanoribbon}
\shortauthors{C. M. V. de Ara\'ujo et al.}

\title[mode=title]{Multiproperty Atomistic Characterization of a
Synthesized Pyrazinic Nitrogen-Doped Porous Armchair Graphene Nanoribbon}

\author[1,2]{Cicera M. V. de Ara\'ujo}
\author[3]{Isaac de M. F\'elix}
\author[1,2]{Willian F. Radel}
\author[4]{Raphael B. de Oliveira}
\author[5]{Guilherme da S. L. Fabris}
\author[5]{Douglas S. Galv\~ao}
\author[1,2,6]{Marcelo L. Pereira Junior}[orcid=0000-0001-9058-510X]
\cormark[1]
\ead{marcelo.lopes@unb.br}

\affiliation[1]{organization={Laboratory of NanoEngineering,
College of Technology, University of Bras\'ilia}, city={Bras\'ilia}, state={DF}, postcode={70910-900}, country={Brazil}}
\affiliation[2]{organization={Graduate Program in Physics, Institute of Physics,
University of Bras\'ilia}, city={Bras\'ilia}, state={DF}, postcode={70910-900}, country={Brazil}}
\affiliation[3]{organization={Center for Agri-Food Science and Technology,
Federal University of Campina Grande}, city={Pombal}, state={PB}, postcode={58840-000}, country={Brazil}}
\affiliation[4]{organization={Institute of Physics,
Federal University of Rio Grande do Norte}, city={Natal}, state={RN}, postcode={59078-970}, country={Brazil}}
\affiliation[5]{organization={Department of Applied Physics,
Gleb Wataghin Institute of Physics, University of Campinas}, city={Campinas}, state={SP}, postcode={13083-859}, country={Brazil}}
\affiliation[6]{organization={Department of Electrical Engineering,
College of Technology, University of Bras\'ilia}, city={Bras\'ilia}, state={DF}, postcode={70910-900}, country={Brazil}}
\cortext[1]{Corresponding author.}

\begin{abstract}
Graphene nanoribbons (GNRs) obtained by on-surface synthesis combine a width-controlled band gap with atomic precision, making chemical decoration of their edges and interiors a practical route to designed electronic and optical behavior. Among the structures already realized, Ullmann coupling of a phenazine-bearing precursor yields a nine-atom-wide armchair ribbon that is periodically perforated and carries two two-coordinated nitrogen atoms at the rim of every pore, a site fixed by the precursor rather than by the reaction. Here we characterize this ribbon across the properties relevant to device operation, alongside the pristine and the undoped porous ribbon, thereby disentangling the distinct roles of pore formation and nitrogen incorporation. We find that perforation primarily governs the mechanical and thermal response, whereas nitrogen substitution predominantly controls the electronic and optical properties. Perforation removes the atomic row that carries the frontier states of the pristine ribbon, opening the gap into the range measured by tunneling spectroscopy while substantially reducing stiffness and thermal conductivity. Substitution leaves the mechanical response essentially unchanged and contributes only a secondary reduction of the conductivity through reweighting of the phonon spectrum. Its primary effect is electronic: it selectively narrows the conduction band, binds the lowest exciton by \SI{414}{\milli\electronvolt}, and displaces the absorption edge into the red. The effective phonon mean free path we obtain, \SI{9.9}{\nano\meter}, matches the length to which the synthesis presently limits these ribbons, placing reported samples at the ballistic-to-diffusive crossover. Adsorption on Au(111), Ag(111), and Cu(111) is energetically nearly indistinguishable and remains van der Waals in character, accounting for the substrate tolerance observed experimentally. Finally, we show that graphitic rather than pyrazinic nitrogen would instead make the ribbon metallic, except when the two atoms occupy pore-rim sites, where a ferromagnetic semiconducting state emerges carrying a magnetic moment of \(2\,\mu_{\mathrm{B}}\) per unit cell.
\end{abstract}


\maketitle

\section{Introduction}\label{sec:intro}

Carbon sustains the same $sp^2$ bonding in three, two and one dimensions. Graphite, graphene, nanotubes and nanoribbons share a local structure and differ only in how it is extended, so that geometry can be varied while the chemistry is held fixed. For this reason, low-dimensional carbon has become a standard setting for relating structure and properties, and devices built on it range from transparent electrodes and gas sensors to thermal management and quantum circuits \citep{Novoselov2004,Geim2007,Paupitz2026}.

Graphene delivered record carrier mobility and the highest measured thermal conductivity of any material, yet it has no band gap, which excludes it from the switching role that electronics requires \citep{CastroNeto2009,Balandin2008}. Confining one of its dimensions recovers that gap through quantum confinement, and the resulting graphene nanoribbons (GNRs) are semiconducting with a gap that scales inversely with width and depends on the edge termination through three distinct families \citep{Nakada1996,Son2006,Son2007erratum}. The problem therefore moves from the material to the fabrication, because a gap that depends on width to within one atomic row is useful only if the width can be controlled to that precision.

Bottom-up synthesis on metal surfaces provides that precision through Ullmann coupling of designed halogenated precursors followed by cyclodehydrogenation, which yields ribbons of controlled geometry. The molecular building block, rather than lithography, sets width, edge and topology \citep{Grill2007,Cai2010,Talirz2016}. The approach now delivers armchair ribbons of chosen width, zigzag edges, chiral edges and heterojunctions, and the substrate itself participates by directing the polymerization \citep{Lafferentz2012,Fan2015,Ruffieux2016,Cai2014}. Nine-atom wide armchair ribbons grown this way have already operated as the channel of a short-channel field-effect transistor, so that the platform is no longer at the stage of proof of principle \citep{Talirz2017,Llinas2017,Wang2021quantum}.

On this basis, the field has accumulated a broad repertoire of carbon ribbons in which the local connectivity, rather than the width alone, becomes the design variable. Ribbons with coronene edges, necklace-like ribbons, chemically porous ribbons, topologically engineered segments and ribbons built from non-hexagonal rings have all been examined. In each case, rearranging bonds within a fixed composition alters charge transport, elastic response and optical absorption \citep{PereiraJr2020coronene,PereiraJr2020necklace,daCunha2021,PereiraJr2021topo,PereiraJr2020chevron,DeSousa2026,Gomes2026}. Two-dimensional analogues follow the same logic, with porous carbon networks and non-benzenoid allotropes tuned by the same principle \citep{Bieri2009,Moreno2018,Pereira2022biphenylene,Lima2025,Alves2025}.

Two further degrees of freedom have been added to width and edge. Periodic porosity opens a gap by confining the $\pi$ system within a lattice of designed voids, and has been proposed for molecular sieving and sensing \citep{Pedersen2008,Jiang2009,Celebi2014}. Heteroatom substitution changes the chemistry of a single site without disturbing the lattice, and nitrogen is the natural choice because it is isoelectronic with carbon and enters graphitic carbon at chemically distinct positions. Individual nitrogen dopants have been imaged atom by atom in two-dimensional graphene, and the same substitution has been carried into ribbons and into amorphous carbon \citep{Zhao2011,Schiros2012,Lin2015,Santos2025}.

Graphitic nitrogen is three-coordinated and donates its fifth electron to the $\pi$ band, shifting the Fermi level and acting as an $n$ dopant, whereas pyridinic and pyrazinic nitrogen are two-coordinated and retain a lone pair in the plane, so that their effect on the $\pi$ system differs qualitatively \citep{Wang2012Ngraphene,Lazar2019,Joucken2015,Wen2023}. The coordination of the site therefore matters more than its concentration. The electronic interaction between two nitrogen atoms also depends on how far apart they sit \citep{Tison2015}. Achieving one site to the exclusion of the others was for a long time the bottleneck of nitrogen doping, and syntheses that target a given coordination often converge to another \citep{Zhang2022,Bassi2024,Wang2018chiralGNR}.

Ullmann coupling of tetra\-bromo\-tetra\-benzo[$a,c,h,j$]\allowbreak phen\-azine, a precursor that carries the pyrazine ring already formed, resolved that selectivity. It yields a nine-atom-wide armchair ribbon, perforated once per unit cell and carrying two pyrazinic nitrogen atoms at the rim of every pore. The coordination is fixed by the precursor rather than by the reaction \citep{Pawlak2020,Wang2017pyrazinePAH}. The ribbon was first grown on Ag(111) and later on Cu(111), Ag(100) and Ag(110). The same backbone, pore periodicity and nitrogen site were recovered on all four, indicating that the molecular design dominates over the symmetry of the support \citep{Ceccatto2026}. Scanning probe measurements established the structure and the transport gap, and the frontier orbitals were assigned a donor-acceptor character.

The synthesis established the geometry and the gap, but several properties required for device operation remain to be determined. These include the mechanical response under load, heat flow along the ribbon, the optical response once electron-hole interaction is included, and the behavior at the interface with the metal on which the ribbon is grown. These properties are also the ones that cannot be read from the structure alone.

In this work, we characterize the synthesized pyrazinic nitrogen-doped porous ribbon (9-NAGNR) across that full set, always alongside the pristine nine-atom-wide armchair ribbon (9-AGNR) and the porous ribbon without nitrogen (9-PAGNR), so that the effect of the pore can be separated from the effect of the nitrogen. We find that the pore is the structural perturbation and the nitrogen is an electronic one. The pore alone accounts for the loss of stiffness and for most of the reduction in thermal conductivity, and it opens the gap enough to move the absorption edge from the infrared into the visible. Nitrogen leaves the mechanics untouched, further reduces the thermal conductivity by reweighting the phonon spectrum, and acts on the conduction band alone, strengthening the electron-hole interaction and shifting the optical onset further to the red. On the three noble metal surfaces, the ribbon physisorbs indistinguishably, which supplies a mechanism for the substrate tolerance seen in synthesis. We close with a theoretical extension that asks what the same ribbon would do if the nitrogen occupied graphitic rather than pyrazinic sites, and find a route to a ferromagnetic semiconductor that the thermodynamics of the synthesis does not favor.

\section{Methodology}\label{sec:metodologia}
Geometry optimizations and electronic structure calculations were performed within the framework of density functional theory (DFT) using the Spanish Initiative for Electronic Simulations with Thousands of Atoms (SIESTA) package, in which the Kohn-Sham wave functions are expanded in a basis set of numerical atomic orbitals \citep{Soler2002,Junquera2001,Garcia2020}. Exchange-correlation effects were described within the generalized gradient approximation (GGA) using the Perdew-Burke-Ernzerhof (PBE) functional \citep{Perdew1996}, while core electrons were represented by norm-conserving Troullier-Martins pseudopotentials \citep{Troullier1991}. A double-$\zeta$ polarized (DZP) basis set was employed, together with a real-space grid defined by a mesh cutoff of \SI{700}{\rydberg}. Brillouin-zone integrations were carried out using a shifted Monkhorst-Pack $10\times1\times1$ $k$-point mesh along the ribbon direction \citep{Monkhorst1976}. To avoid interactions between periodic images, a vacuum region of \SI{40}{\angstrom} was introduced along the two non-periodic directions. Structural relaxations, including both atomic coordinates and lattice parameters, were performed using the conjugate-gradient algorithm until the residual forces on each atom and the residual stress were smaller than \SI{1e-3}{\electronvolt\per\angstrom} and \SI{0.05}{\giga\pascal}, respectively.

Semilocal functionals underestimate the band gap of graphene nanoribbons by an amount that grows as the frontier states localize, which quasiparticle calculations have quantified at between \num{0.5} and \SI{3.0}{\electronvolt} \citep{Yang2007QP}. The gaps reported here were therefore recomputed with the screened hybrid functional of Heyd, Scuseria and Ernzerhof (HSE06) \citep{Heyd2003,Heyd2006erratum,Krukau2006} on the relaxed geometries, using the exact-exchange implementation for numerical atomic orbitals of HONPAS, the Hefei Order-N Packages for Ab initio Simulations \citep{Qin2015,Shang2011} with the same basis, mesh and $k$ sampling.

Phonon dispersions were obtained by finite displacements in a $2\times1\times1$ supercell, with the force constants and the dynamical matrix assembled by phonopy \citep{Parlinski1997,Togo2015,Togo2023}. Thermal stability was probed by \textit{ab initio} molecular dynamics (AIMD) in the canonical ensemble with a Nos\'e thermostat, using a \SI{1}{\femto\second} time step over \SI{5}{\pico\second} at \SI{1500}{\kelvin} \citep{Nose1984}.

The mechanical response was obtained by straining the cell uniaxially along the ribbon axis in steps, relaxing the atomic positions at each step. A ribbon has no unambiguous thickness, so the stress is reported per unit width in \si{\newton\per\meter}, the two-dimensional convention used for graphene and other single-layer carbon structures, which removes dependence on an assumed thickness.

Lattice thermal conductivity was calculated using reverse non-equilibrium molecular dynamics (RNEMD) simulations implemented in the Large-scale Atomic/Molecular Massively Parallel Simulator (LAMMPS) package \citep{Plimpton1995,Thompson2022}. The heat flux was imposed through the Müller-Plathe algorithm, which exchanges kinetic energy between designated hot and cold regions of the system, while the resulting temperature gradient is monitored along the transport direction \citep{MullerPlathe1997}. Kinetic energy swaps were performed every 500 simulation timesteps. Interatomic interactions were described by the reactive force field ReaxFF \citep{vanDuin2001,fthenakis2022evaluating}, using a parameter set developed for nitrogen-containing carbon structures \citep{Kowalik2019}. This parameterization allows all three nanoribbon models to be treated within a single force-field framework. The equations of motion were integrated with a time step of \SI{0.1}{\femto\second}, and each simulation was run for $20\times10^6$ timesteps, corresponding to a total simulation time of \SI{2}{\nano\second}. Ribbon lengths ranged from \SI{5}{\nano\meter} to \SI{187}{\nano\meter} for the pristine nanoribbon and up to \SI{94}{\nano\meter} for the porous nanoribbons. A thickness of \SI{3.35}{\angstrom} was assumed for all nanoribbons. The thermal conductivity was then obtained from Fourier's law using the imposed heat flux and the steady-state temperature gradient.

The optical response was obtained from a tight-binding representation built directly from the converged HONPAS Hamiltonian in the numerical atomic orbital basis, extracted with sisl \citep{sisl} and passed to a Wannier-based tight-binding code for excitonic and optoelectronic properties (WanTiBEXOS) \citep{Dias2023}. Spectra were obtained within the independent-particle approximation (IPA) and, including the electron-hole interaction, by solving the Bethe-Salpeter equation (BSE) \citep{Salpeter1951,Rohlfing2000}, following the procedure adopted previously for carbon and MXene monolayers \citep{Gomes2025,Aparicio2025acsaem,Aparicio2025nanoscale}. Light polarized along the ribbon axis dominates the response in one dimension, the transverse component being suppressed by depolarization.

The ribbon-on-metal systems were modeled using a three-layer (111) slab of Au, Ag, or Cu. To achieve commensurability with each substrate, the ribbon was strained by less than \SI{1.5}{\percent} and kept rigid while the interface separation was varied. The isolated-ribbon energy in Eq.~\eqref{eq:eadh} was evaluated in the same strained cell adopted for the interface calculations, ensuring cancellation of the strain contribution to the adhesion energy. Dispersion interactions were described using the van der Waals density functional (vdW-DF) \citep{Dion2004,Dion2005erratum,RomanPerez2009}, together with the same basis set and a $6\times3\times1$ $k$-point grid. This approach follows the rigid single-point protocol previously employed for layered heterostructures \citep{Santos2021}.

Spin polarization was considered for all systems. Ferromagnetic and antiferromagnetic configurations were initialized independently and compared with the corresponding non-magnetic solution on the same geometry. Magnetic ground states were obtained only for the systems discussed below.

\section{Results and Discussion}\label{sec:resultados}

\begin{figure*}[pos = !t]
  \centering
  \includegraphics[width=0.9\linewidth]{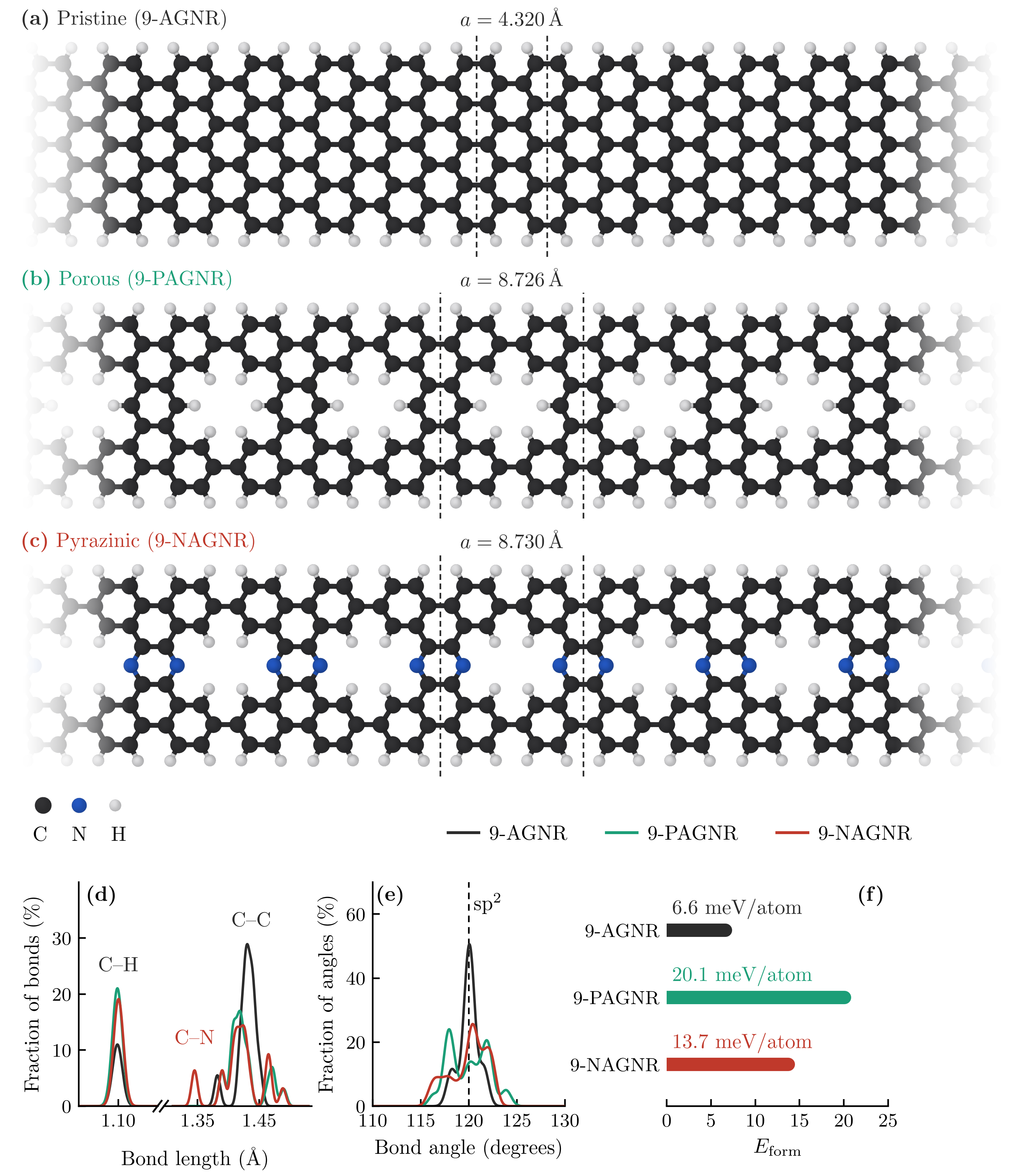}
  \caption{(a-c) The three ribbons studied here, drawn to the same scale and fading
at both ends to indicate periodicity, with two vertical lines marking one unit cell. (d,e) Distributions of bond lengths and bond angles. (f) Formation energy per atom.}
  \label{fig:estrutura}
\end{figure*}

Because the electronic, mechanical, and thermal properties of the nanoribbons are rooted in their atomic structure, the comparison begins with an analysis of their geometries. Figure~\ref{fig:estrutura}(a-c) shows the optimized structures on a common scale, while panels (d) and (e) summarize the bond-length and bond-angle distributions. The relative stability of the ribbons is assessed through the formation energies presented in panel (f).

The pristine ribbon has a lattice parameter of \SI{4.32}{\angstrom} and the composition C$_{18}$H$_4$. Removing one benzene ring per two pristine cells and terminating the resulting dangling bonds gives the porous ribbon, C$_{30}$H$_{14}$, with \SI{8.726}{\angstrom}. Replacing two of the pore-edge C-H groups by nitrogen gives the pyrazinic ribbon, C$_{28}$H$_{12}$N$_2$, with \SI{8.730}{\angstrom}. The pore therefore dilates the lattice by \SI{0.99}{\percent} with respect to two pristine cells, while the nitrogen leaves it unchanged to within \SI{0.05}{\percent}. The computed value for the doped ribbon lies \SI{0.03}{\angstrom} above the range of \SIrange{8.4}{8.7}{\angstrom} measured across four substrates \citep{Pawlak2020,Ceccatto2026}, the slight overestimate expected from a gradient-corrected functional.

Each nitrogen sits at the rim of the pore with two carbon neighbors rather than the three it would have inside the ring, and the four resulting C-N bonds measure \SI{1.35}{\angstrom} with no spread among them, the four being equivalent by symmetry. That length is characteristic of a C=N double bond and is \SI{0.06}{\angstrom} shorter than the C-N distance obtained when the same lattice carries graphitic nitrogen, which sits inside the ring with three carbon neighbors instead of two at the pore edge. The short bond pulls the pore edges inward, narrowing the carbon frame from \SI{9.92}{\angstrom} in the porous ribbon to \SI{9.70}{\angstrom} in the doped one. The angle centered on the nitrogen closes to \ang{118.2}, the compression expected when a lone pair occupies an in-plane $sp^2$ orbital rather than a bond, as it does in pyridine and pyrazine. A two-coordinated nitrogen of this kind contributes one electron to the $\pi$ system, exactly as the carbon it replaced, and keeps the remaining pair in the plane.

The bond-length distributions in Figure~\ref{fig:estrutura}(d) show that it is the pore, not the nitrogen, that reorganizes the lattice. The mean carbon-carbon distance is \SI{1.43}{\angstrom} in all three ribbons, but its spread grows by two thirds when the pore is opened, from \SI{0.016}{\angstrom} to \SI{0.027}{\angstrom}. Short bonds lie within benzene-like units and long ones, up to \SI{1.49}{\angstrom}, on the biphenyl-like bridges that connect them. In chemical terms the porous ribbons are polymers of weakly coupled rings, whereas the pristine ribbon is confined graphene, and this distinction recurs below in the flattened frontier bands, in the shorter effective phonon mean free path and in the stronger electron-hole binding. The angular distributions in \ref{fig:estrutura}(e) broaden in the same way while remaining centered on \ang{120}, and no angle falls below \ang{116}, so the $sp^2$ character and the planarity survive intact in all three systems.

The formation energy per atom in Figure~\ref{fig:estrutura}(f) follows from the total energy and the chemical potentials of the reservoirs,
\begin{equation}
E_\mathrm{form} = \frac{1}{N}\left(E_\mathrm{tot} - \sum_i n_i\,\mu_i\right),
\label{eq:eform}
\end{equation}
where $n_i$ and $\mu_i$ are the number of atoms and the chemical potential of species $i$, and $N$ is the total number of atoms. The reservoirs are graphite, molecular hydrogen and molecular nitrogen. Opening the pore costs \SI{13.5}{\milli\electronvolt} per atom relative to the pristine ribbon. It creates six new edge C-H groups and interrupts the extended aromaticity. Substituting two of those groups by nitrogen recovers \SI{6.4}{\milli\electronvolt} per atom, nearly half the cost. The gain comes from the aromatic stabilization of the pyrazine ring and from removing two high-energy edge terminations. The doped ribbon is thus more stable than the porous ribbon of the same topology, and all three values remain below \SI{21}{\milli\electronvolt} per atom, so that the synthetic route, rather than the thermodynamics, selects which structures are accessible.

\begin{figure*}[pos = !t]
  \centering
  \includegraphics[width=\linewidth]{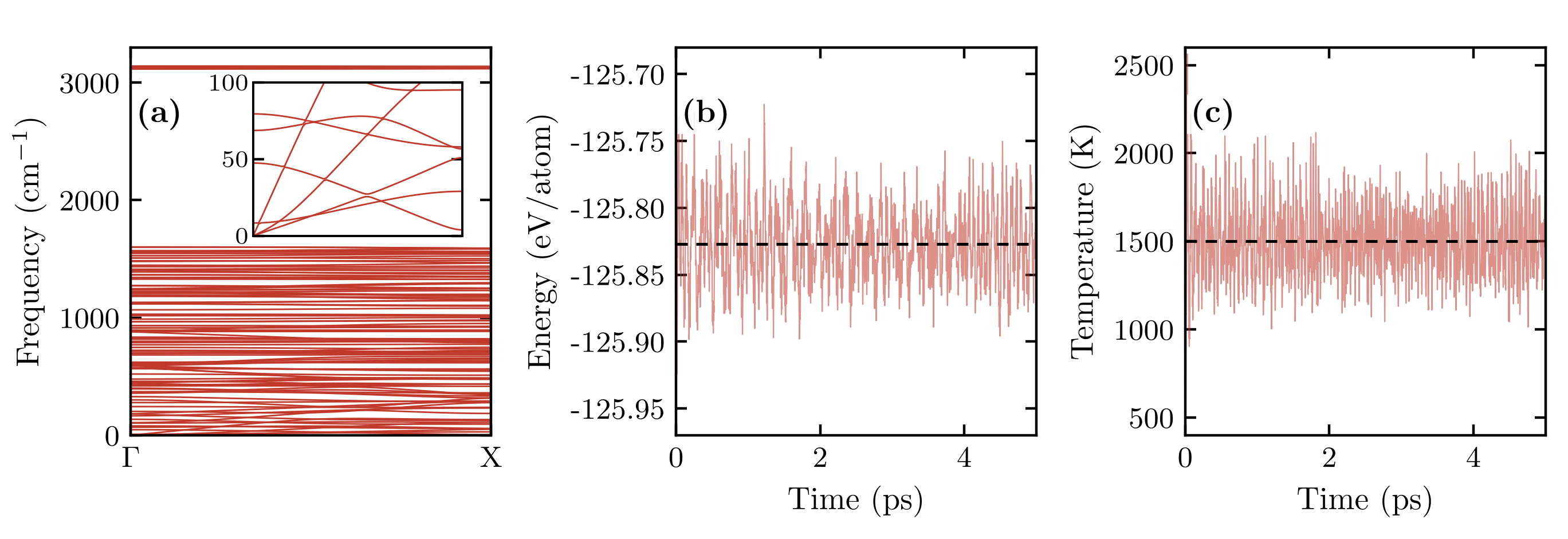}
  \caption{Dynamic and thermal stability of the 9-NAGNR. (a) Phonon dispersion along $\Gamma$-X. (b,c) Kohn-Sham energy per atom and instantaneous temperature along 5\,ps of AIMD at \SI{1500}{\kelvin}.}
  \label{fig:estabilidade}
\end{figure*}

When kinetic factors, rather than thermodynamic stability, govern structure formation, the stability of the resulting ribbon as an isolated system becomes a separate issue. A low formation energy indicates that a structure is energetically accessible, but provides no information about the local curvature of the potential-energy surface or its dynamical stability. This question is addressed for the doped ribbon, the system of primary interest in the present work, through the phonon spectrum shown in Figure~\ref{fig:estabilidade}.

As shown in Figure~\ref{fig:estabilidade}(a), all phonon frequencies are positive, extending up to \SI[per-mode=power]{3141}{\per\centi\meter}, while the four acoustic branches emerging from the $\Gamma$ point increase monotonically without exhibiting imaginary frequencies. These features demonstrate the dynamical stability of the ribbon. The highest-frequency branch is separated from the remaining modes by a gap of approximately \SI[per-mode=power]{1500}{\per\centi\meter} and corresponds to C--H stretching vibrations, which are effectively decoupled from the rest of the spectrum because of the large mass mismatch between carbon and hydrogen. Projection of the phonon eigenvectors onto the atomic species reveals that nitrogen contributions are concentrated in the low- and intermediate-frequency regions. Modes for which nitrogen accounts for more than \SI{10}{\percent} of the total displacement are found below \SI[per-mode=power]{1560}{\per\centi\meter}, predominantly between \SI[per-mode=power]{330}{\per\centi\meter} and \SI[per-mode=power]{710}{\per\centi\meter}, together with an isolated mode near \SI[per-mode=power]{240}{\per\centi\meter} that exhibits the largest nitrogen participation, reaching \SI{45}{\percent}. Because nitrogen is much heavier than hydrogen and has a mass comparable to that of carbon, its motion is primarily associated with ring deformations rather than with the high-frequency C--H stretching modes that dominate the upper part of the spectrum.

The thermal stability of the doped ribbon was further assessed by ab initio molecular dynamics (AIMD) at \SI{1500}{\kelvin}. As shown in Figure~\ref{fig:estabilidade}(c), the instantaneous temperature fluctuates around an average value of \SI{1500.4}{\kelvin}, with a standard deviation of \SI{204.8}{\kelvin}, close to the \SI{189}{\kelvin} expected from equipartition for a simulation cell containing \num{42} atoms. This agreement indicates that the trajectory remains properly thermalized throughout the simulation. Over the \SI{5}{\pico\second} trajectory, the Kohn-Sham energy shown in Figure~\ref{fig:estabilidade}(b) exhibits a drift of only \SI{-0.4}{\milli\electronvolt} per atom per picosecond. For comparison, the breaking of a single C--C bond in a cell of this size would release approximately \SI{110}{\milli\electronvolt} per atom, corresponding to roughly four times the standard deviation of the energy fluctuations. No event of this magnitude is observed, and the atomic connectivity remains unchanged throughout the simulation. The same protocol applied to the pristine and porous ribbons likewise preserves their bonding topology. These results indicate that all three ribbons remain structurally intact even at a temperature approximately five times higher than a typical operating condition.

\begin{figure}[pos = h!]
  \centering
  \includegraphics[width=\linewidth]{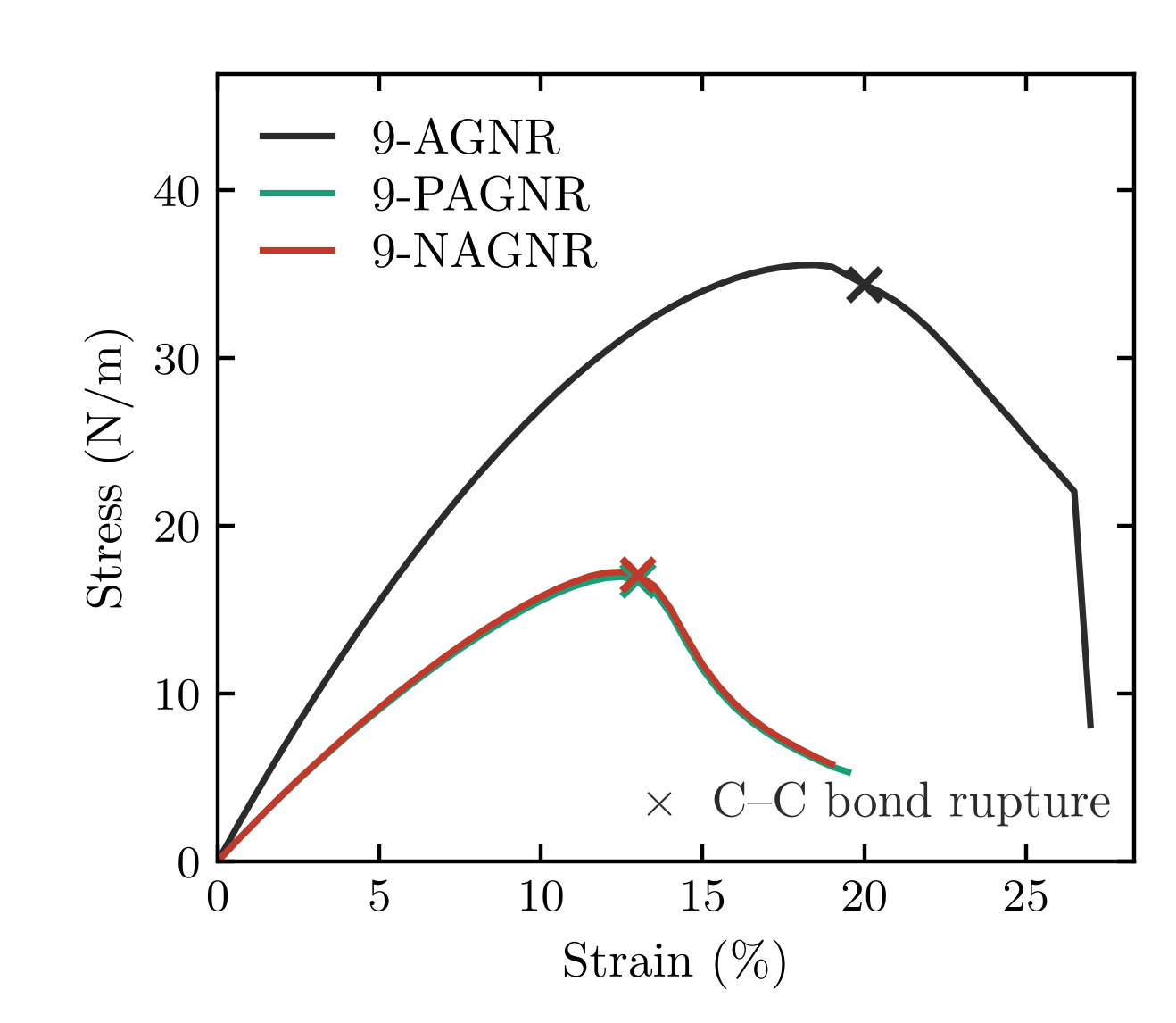}
  \caption{Stress-strain response under uniaxial tension along the ribbon
axis. The curve of the porous ribbon lies under that of the doped one. Crosses mark the strain at which the first carbon-carbon bond breaks.}
  \label{fig:mecanica}
\end{figure}

Thermal stability alone, however, does not guarantee mechanical robustness. Introducing pores reduces the number of load-bearing bonds, and any enhancement in functionality must therefore be balanced against a potential loss of stiffness. To quantify this effect, Figure~\ref{fig:mecanica} presents the stress-strain response of the three ribbons under uniaxial tension applied along the ribbon axis.

The pristine ribbon has a two-dimensional Young modulus of \SI{337.6}{\newton\per\meter}, essentially that of graphene \citep{Lee2008}. It reaches a maximum stress of \SI{35.5}{\newton\per\meter} before the first bond breaks at \SI{20}{\percent} strain. Opening the pore lowers the modulus to \SI{199.4}{\newton\per\meter} and the maximum stress to \SI{17.0}{\newton\per\meter}. Failure occurs at \SI{13}{\percent}. The linear density of carbon-carbon bonds falls by a quarter upon perforation, from \num{25} bonds per \SI{4.32}{\angstrom} to \num{38} per \SI{8.726}{\angstrom}, while the modulus falls by \SI{41}{\percent}, and the discrepancy arises because the remaining bonds are not all aligned with the loading axis. Tension is carried by the two lateral wings of the ribbon and by the biphenyl-like bridges, which work partly in bending. The narrow bridges concentrate the stress that eventually breaks them. This is the same connectivity argument that governs failure in porous carbon membranes and in non-benzenoid networks \citep{Pereira2020NPG,Pereira2022biphenylene,Gomes2026}. The values reported here are ideal strengths of a homogeneous cell \citep{Liu2007}, and a ribbon carrying defects would fail earlier.

Nitrogen alters none of this, and the doped ribbon has a modulus of \SI{200.5}{\newton\per\meter} against \SI{199.4}{\newton\per\meter} for the porous one, a difference of \SI{0.6}{\percent}, with a maximum stress of \SI{17.3}{\newton\per\meter} against \SI{17.0}{\newton\per\meter} and the same rupture strain and failure mode. The origin is geometric, since the two nitrogen atoms sit at the pore rim with their short C-N bonds transverse to the loading axis, so they are not part of the load path, and the bonds that break are carbon-carbon bonds on the bridges. Mechanically the substitution is invisible, so that the topology alone sets the mechanical response.

\begin{figure}[pos = h!]
  \centering
  \includegraphics[width=\linewidth]{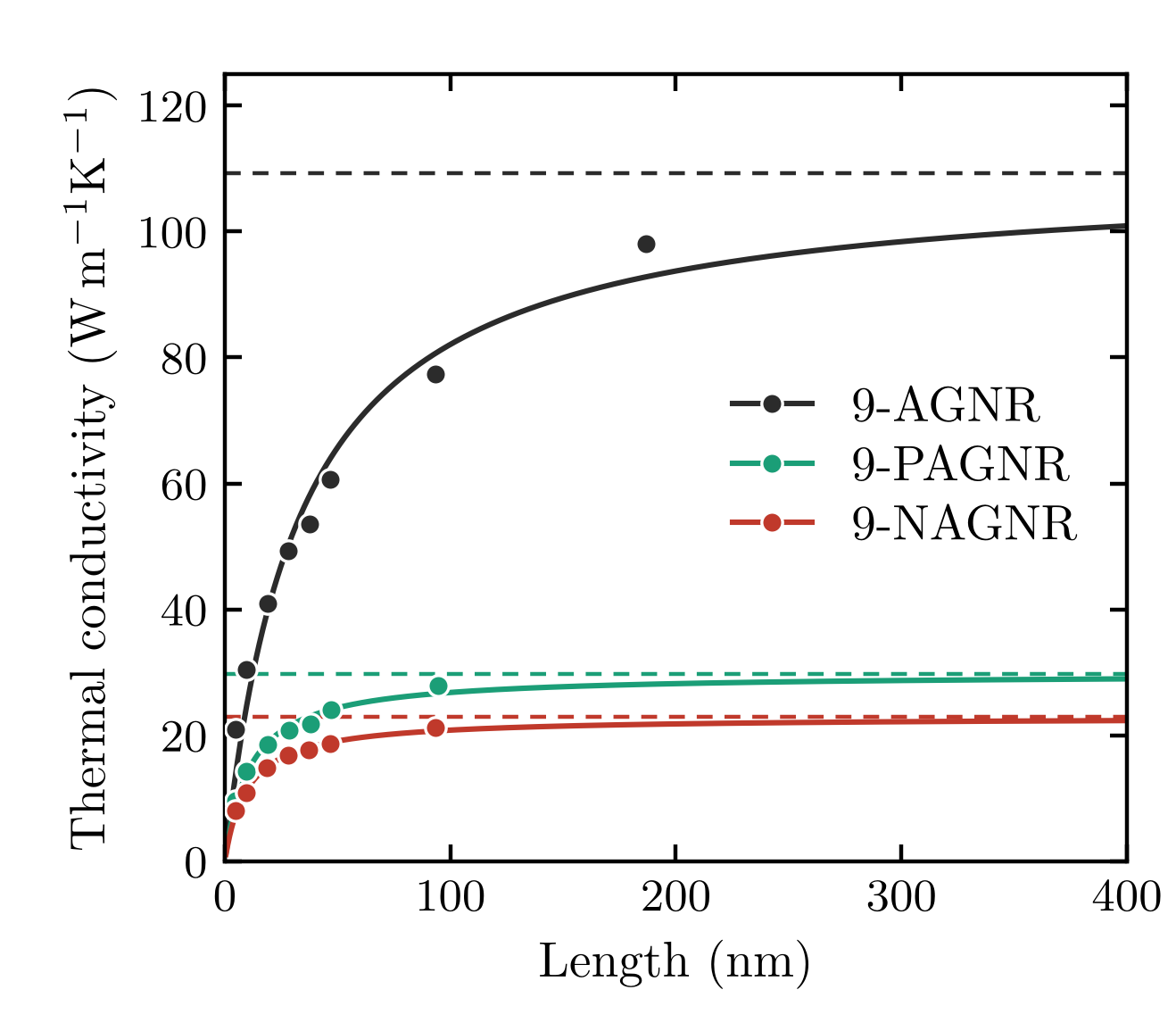}
  \caption{Thermal conductivity against ribbon length from reverse
non-equilibrium molecular dynamics. Solid curves are fits of Eq.~\eqref{eq:kappa} and dashed lines mark the extrapolated plateau.}
  \label{fig:termica}
\end{figure}

Heat removal is the limiting factor in a nanoscale device \citep{Balandin2011}, and the same perforation that cuts bonding paths cuts phonon paths. In contrast with the mechanical response, the nitrogen contributes measurably here, and the two effects can be separated. Figure~\ref{fig:termica} shows the thermal conductivity as a function of ribbon length.

The conductivity grows with length and saturates, following the form obtained when the intrinsic and boundary resistances add,
\begin{equation}
\kappa(L) = \frac{\kappa_\infty}{1 + \Lambda_\text{eff}/L},
\label{eq:kappa}
\end{equation}
so that $1/\kappa$ is linear in $1/L$, the intercept gives the converged conductivity $\kappa_\infty$ and the slope gives $\Lambda_\text{eff}/\kappa_\infty$, from which the effective phonon mean free path $\Lambda_\text{eff}$ follows \citep{Schelling2002}. At $L=\Lambda_\text{eff}$ the conductivity is half its limit. 
Here, $\Lambda_{\text{eff}}$ is understood as an effective transport length scale obtained from the length-dependent thermal conductivity, rather than as a unique
microscopic mean free path associated with a particular phonon mode.
Fitting the simulated points without weighting gives $\kappa_\infty = \SI{109.3\pm7.3}{\watt\per\meter\per\kelvin}$ with $\Lambda_\text{eff} = \SI{33.1}{\nano\meter}$ for the pristine ribbon, in agreement with previous reports \cite{liao2011thermally,ng2012molecular}. The porous ribbon gives \SI{29.8\pm1.0}{\watt\per\meter\per\kelvin} with \SI{11.0}{\nano\meter}, and the doped one \SI{22.9\pm0.5}{\watt\per\meter\per\kelvin} with \SI{9.9}{\nano\meter}. Lateral confinement alone already brings the pristine ribbon a factor of roughly fifty below graphene \citep{Balandin2008,Hu2009,Evans2010}, and the pore removes a further factor of \num{3.7}.

From the three-dimensional isotropic kinetic approximation, $\kappa = \tfrac{1}{3} C v \Lambda$, the ratio $\kappa_\infty/\Lambda_\text{eff}$ measures the product of heat capacity and group velocity, and it falls only from \num{3.30} to \SI{2.70}{\watt\per\meter\per\kelvin\per\nano\meter} once the pore is present, while $\Lambda_\text{eff}$ falls by a factor of three. The pore therefore suppresses conduction mainly by creating internal surface that scatters phonons, and only secondarily by softening the spectrum, in agreement with what has been established for graphene nanomeshes and for defect-engineered ribbons \citep{Feng2016,Yousefi2020,Haskins2011}.

Nitrogen removes a further \SI{23}{\percent}, and it reverses the weight of the two factors, since the effective phonon mean free path falls by only \SI{10}{\percent} whereas the capacity-velocity product falls by \SI{15}{\percent}. Nitrogen is \SI{17}{\percent} heavier than carbon, and because the substitution is periodic it shifts the phonon spectrum rather than introducing scattering centers, which is why the effective mean free path is the quantity that barely moves. The same distinction separates ordered substitution from the isotopic disorder whose scattering has been measured directly in graphene \citep{Chen2012}. The short and stiff C=N bond adds a local force-constant mismatch at the pore rim. This is the first property in which the substitution is resolved beyond the effect of the pore, since thermal transport responds to the mass and force-constant changes the substitution makes to the spectrum whereas the elastic response does not.

This has a direct consequence for measurement, because the effective phonon mean free path of the doped ribbon, \SI{9.9}{\nano\meter}, matches the lengths of about \SI{10}{\nano\meter} to which the adatom-assisted route currently limits these ribbons \citep{Pawlak2020}. Evaluating Eq.~\eqref{eq:kappa} at that length gives \SI{11.5}{\watt\per\meter\per\kelvin}, so a ribbon of the size now synthesized conducts at half the rate its own material would. The contrast between the pristine and the doped ribbon, a factor of \num{4.8} in the long ribbon limit, narrows to a factor of \num{2.2} at that length. Length-dependent conductivity of this kind has been measured directly in suspended graphene \citep{Xu2014}, so that a thermal measurement on these ribbons characterizes the specimen as much as the material.

\begin{figure*}[pos = !t]
  \centering
  \includegraphics[width=\linewidth]{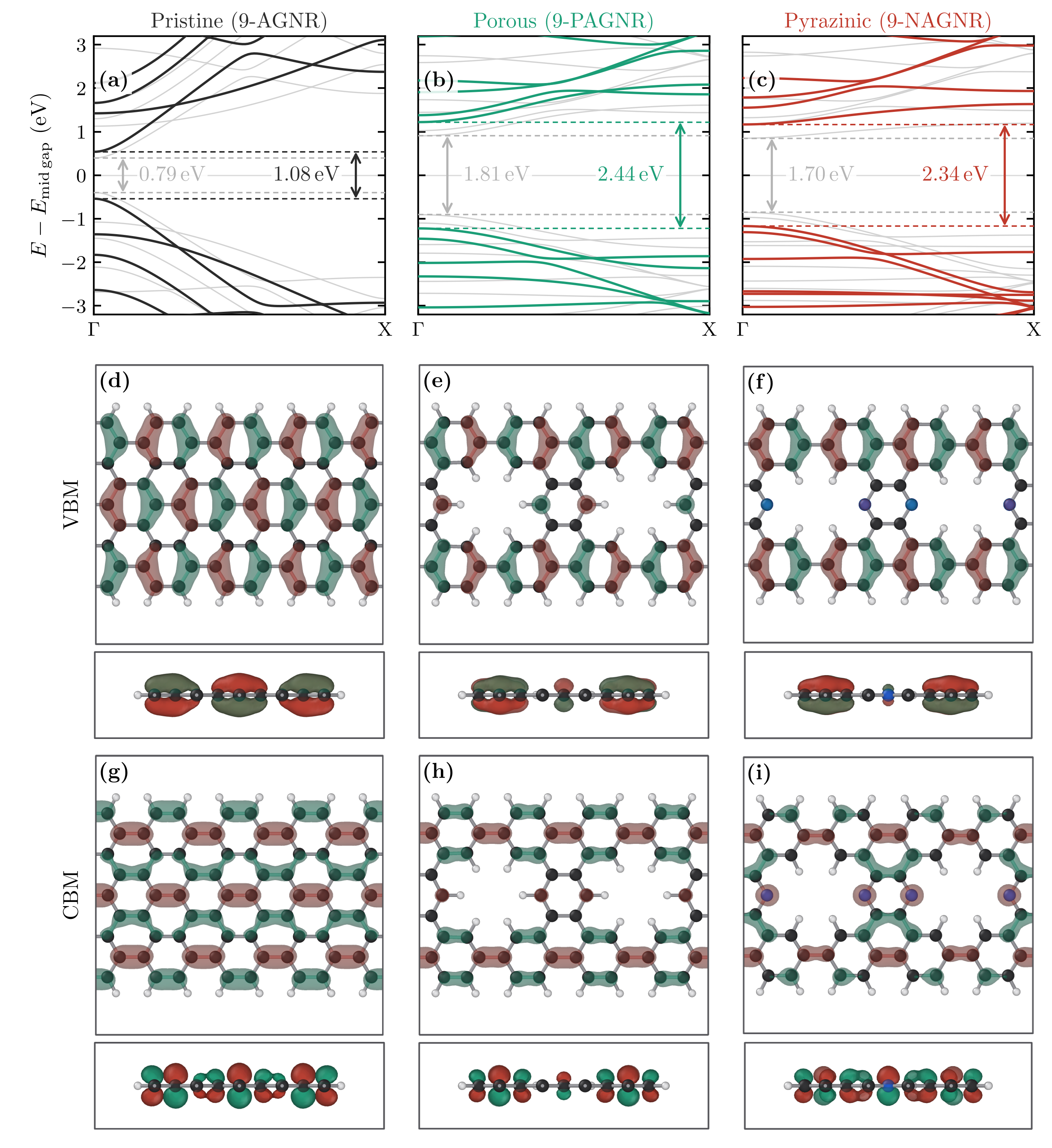}
  \caption{Band structures and frontier orbitals. (a-c) PBE and HSE06 bands,
aligned at mid-gap. (d-f) Valence band maximum and (g-i) conduction band minimum, shown from above and in lateral cut at a common isovalue.}
  \label{fig:eletronica}
\end{figure*}

The elastic response follows from the lattice alone and the thermal one almost entirely so. Its effect appears in the electronic structure, and the band gap is also the quantity the synthesis determined, which makes it the natural point of comparison with experiment. Figure~\ref{fig:eletronica}(a-c) shows the band structures of the three ribbons at both levels of theory, and \ref{fig:eletronica}(d-i) the frontier orbitals in real space.

All three ribbons have a direct gap at the zone center. At the semilocal level the gaps are \SI{0.79}{\electronvolt}, \SI{1.81}{\electronvolt} and \SI{1.70}{\electronvolt}, and the hybrid functional raises them to \SI{1.08}{\electronvolt}, \SI{2.44}{\electronvolt} and \SI{2.34}{\electronvolt}. Scanning tunneling spectroscopy (STS) on the doped ribbon grown on Ag(111) locates the occupied and empty frontier resonances \SI{2.7}{\electronvolt} apart, while the same work quotes \SI{2.2}{\electronvolt} in its closing summary \citep{Pawlak2020}. The hybrid result lies above the \SI{2.2}{\electronvolt} quoted as the gap and below the \SI{2.7}{\electronvolt} separation between resonance maxima, which exceeds a gap by the width of the resonances themselves, whereas the semilocal result lies below both. Perforation and nitrogen together open the gap by \SI{1.26}{\electronvolt} relative to the pristine ribbon, in close agreement with the \SI{1.3}{\electronvolt} inferred experimentally for the same pair \citep{Pawlak2020}. Perforation alone accounts for \SI{1.36}{\electronvolt} of that opening and the nitrogen returns \SI{0.10}{\electronvolt} of it. An opening of the same kind, \SI{2.17}{\electronvolt}, follows when pores are introduced into a twelve-atom wide ribbon \citep{Fan2024,Gomes2025}. Because the ribbon is measured on a metal, where image-charge screening renormalizes the gap downward \citep{Ruffieux2012,Kharche2016,MerinoDiez2017}, a free-standing calculation is expected to lie above a supported measurement, which is the direction of the difference found here.

In the pristine ribbon both the valence band maximum in Figure~\ref{fig:eletronica}(d) and the conduction band minimum in \ref{fig:eletronica}(g) have their largest amplitude on the two carbon atoms of the central row, and the pore removes precisely that row. The perforation therefore excises the frontier states rather than perturbing them uniformly, as the redistributed frontier states of the porous ribbon in \ref{fig:eletronica}(e) and \ref{fig:eletronica}(h) show, and the consequences propagate through the remaining properties. The frontier bandwidth collapses from \SI{2.43}{\electronvolt} to \SI{0.68}{\electronvolt} in the valence band. Through the tight-binding relation $W = 4t$ this corresponds to an intercell coupling of \SI{0.17}{\electronvolt}, against \SI{0.61}{\electronvolt} in the pristine ribbon, so that the porous ribbons behave electronically as chains of weakly coupled molecular units. The same picture accounts for the flattened bands of related quasi-one-dimensional carbon chains \citep{Lage2026}. The hybrid correction accordingly doubles from \SI{0.29}{\electronvolt} to \SI{0.63}{\electronvolt} upon perforation, since exact exchange responds to the degree of localization.

Going from the porous to the doped ribbon narrows the valence bandwidth by \SI{5}{\percent} and the conduction bandwidth by \SI{34}{\percent}, so the nitrogen acts almost exclusively on the conduction band. The two nitrogen atoms carry \SI{4}{\percent} of the valence band maximum against \SI{29}{\percent} of the conduction band minimum, the contrast between the isosurfaces in \ref{fig:eletronica}(f) and \ref{fig:eletronica}(i). Taking the per-atom density in each band relative to the largest in that band, the pore-edge site holds \num{0.79} in the valence band and \num{0.32} in the conduction band while it is carbon; once it becomes nitrogen it falls to \num{0.32} in the valence band and rises to \num{1.00}, the largest conduction amplitude in the ribbon. The origin is the higher electronegativity of nitrogen, which lowers the local $p_z$ level and thereby attracts the antibonding combination rather than the bonding one. The hybrid correction is \SI{0.63}{\electronvolt} in both porous ribbons, unchanged by the substitution, confirming that the nitrogen acts at the single-particle level without altering the degree of localization. Pyrazinic nitrogen therefore behaves as a localized $\pi$ acceptor rather than as a dopant, in contrast with graphitic nitrogen in two-dimensional graphene, which donates an electron and shifts the Fermi level \citep{Zhao2011,Schiros2012,Wen2023}. The valence state on the carbon wings and the conduction state on the nitrogen are spatially separated, which corresponds to the donor-acceptor character assigned experimentally to the frontier orbitals \citep{Pawlak2020}.

\begin{figure*}[pos = !t]
  \centering
  \includegraphics[width=\linewidth]{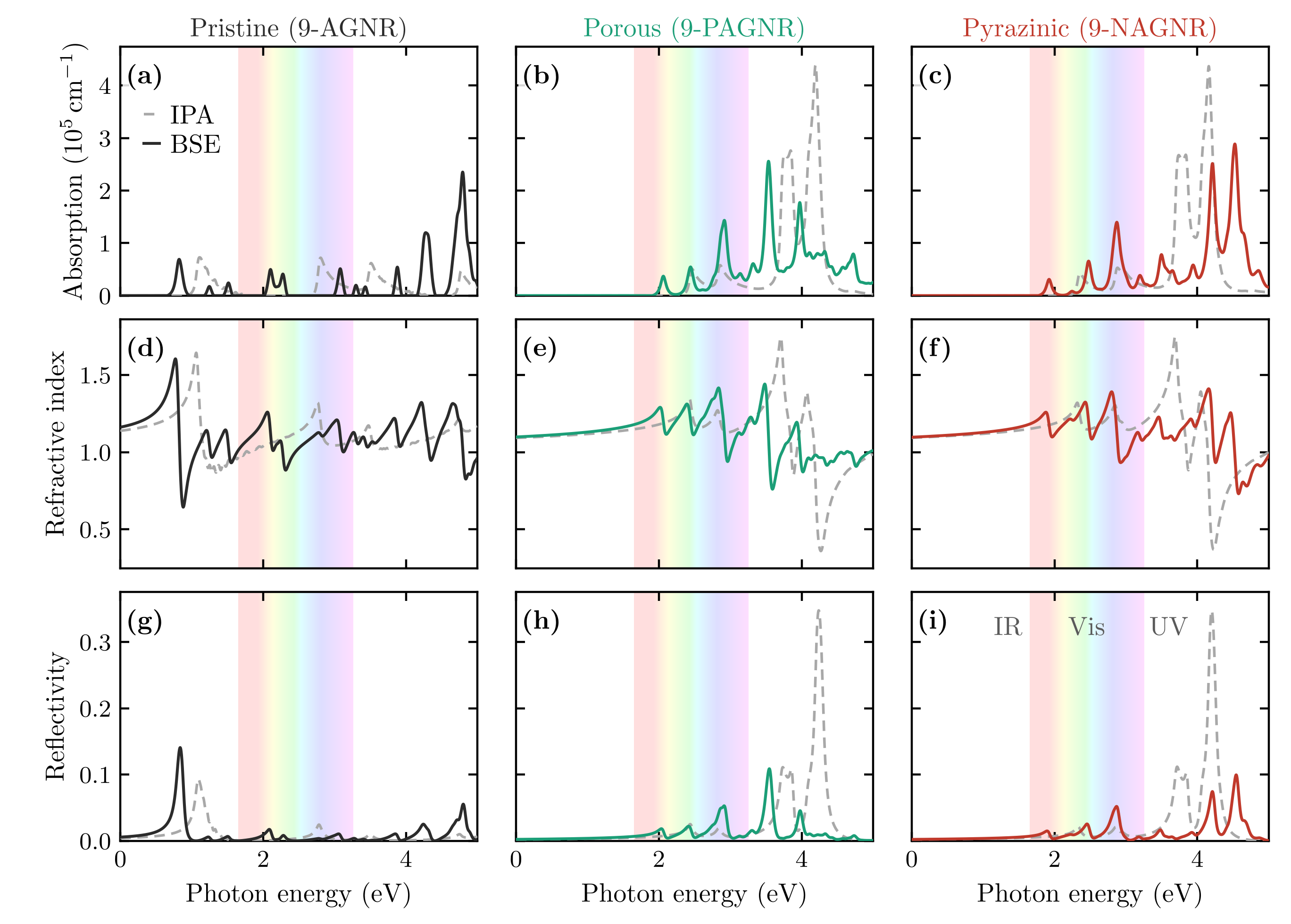}
  \caption{Linear optical response for light polarized along the ribbon axis,
within the IPA and from the BSE. (a-c) Absorption coefficient, (d-f) refractive index and (g-i) reflectivity, in the order pristine, porous, pyrazinic. The shaded band marks the visible range.}
  \label{fig:optica}
\end{figure*}

A gap that has moved into the visible only translates into absorption if the oscillator strength follows, and in a one-dimensional system the electron-hole interaction is too large to be left out \citep{Yang2007Exciton,Prezzi2008}. Figure~\ref{fig:optica}(a-c) shows the absorption coefficient, \ref{fig:optica}(d-f) the refractive index and \ref{fig:optica}(g-i) the reflectivity, each with and without that interaction.

Including electron-hole interactions lowers the absorption edge and redistributes oscillator strength toward the lowest-energy resonances, demonstrating that the optical spectrum cannot be described as a rigidly shifted independent-particle spectrum. The binding energy of the first bright exciton, defined as the difference between the quasiparticle gap and the lowest bright excitation,
\begin{equation}
E_b = E_\mathrm{gap} - E_1, 
\label{eq:eb} 
\end{equation}
is \SI{270}{\milli\electronvolt} for the pristine ribbon, \SI{384}{\milli\electronvolt} for the porous ribbon, and \SI{414}{\milli\electronvolt} for the doped ribbon. This trend follows the progressive flattening of the frontier bands, since a larger reduced effective mass enhances electron-hole confinement. The calculated values lie within the range reported for excitons in carbon nanotubes and narrow armchair graphene nanoribbons, where binding energies of several hundred millielectronvolts have been both predicted and experimentally observed \citep{Wang2005,Denk2014,Tries2020,Gomes2025}. In all cases, the exciton binding energy exceeds the thermal energy at room temperature by roughly one order of magnitude, indicating that the optical response is governed by bound excitonic states rather than free charge carriers. Consequently, the optical and transport gaps remain distinct quantities separated by the exciton binding energy.

The stronger exciton binding found in the doped ribbon, despite its smaller quasiparticle gap, originates from the electronic asymmetry induced by nitrogen incorporation. A reduction of approximately one-third in the conduction-band dispersion increases the effective electron mass, while the localization of \SI{29}{\percent} of the conduction-band-edge state on only two atomic sites gives the excitation a partial charge-transfer character. The hole, in contrast, remains distributed over the carbon backbone. Such spatial asymmetry favors electron-hole attraction and produces a more strongly bound exciton than a conventional Wannier-type state \citep{Knupfer2003}. As a result, the first bright transition is red-shifted from \SI{602}{\nano\meter} in the porous ribbon to \SI{646}{\nano\meter} in the doped ribbon. By contrast, the pristine ribbon exhibits its first intense absorption peak at \SI{1536}{\nano\meter}, in the infrared, and its average absorption across the visible spectrum is nearly one order of magnitude lower than that of the porous structures.

The refractive index and reflectivity follow the same overall trend, as both quantities are linked to the absorption spectrum through the Kramers-Kronig relations. Their main maxima occur below \SI{1}{\electronvolt} for the pristine ribbon and above \SI{3}{\electronvolt} for the porous and doped systems. Averaged across the visible range, the refractive index increases from \num{1.08} in the pristine ribbon to \num{1.20} and \num{1.17} in the porous and doped ribbons, respectively, while the corresponding reflectivities rise from \SI{0.4}{\percent} to \SI{1.4}{\percent} and \SI{1.2}{\percent}. Porosity combined with nitrogen incorporation therefore transforms a weakly absorbing infrared material into an efficient visible-light absorber, consistent with the influence of ring connectivity on the optical response of experimentally realized carbon nanoribbon networks \citep{Gomes2026,Bessa2025}.

\begin{figure*}[pos = !t]
  \centering
  \includegraphics[width=\linewidth]{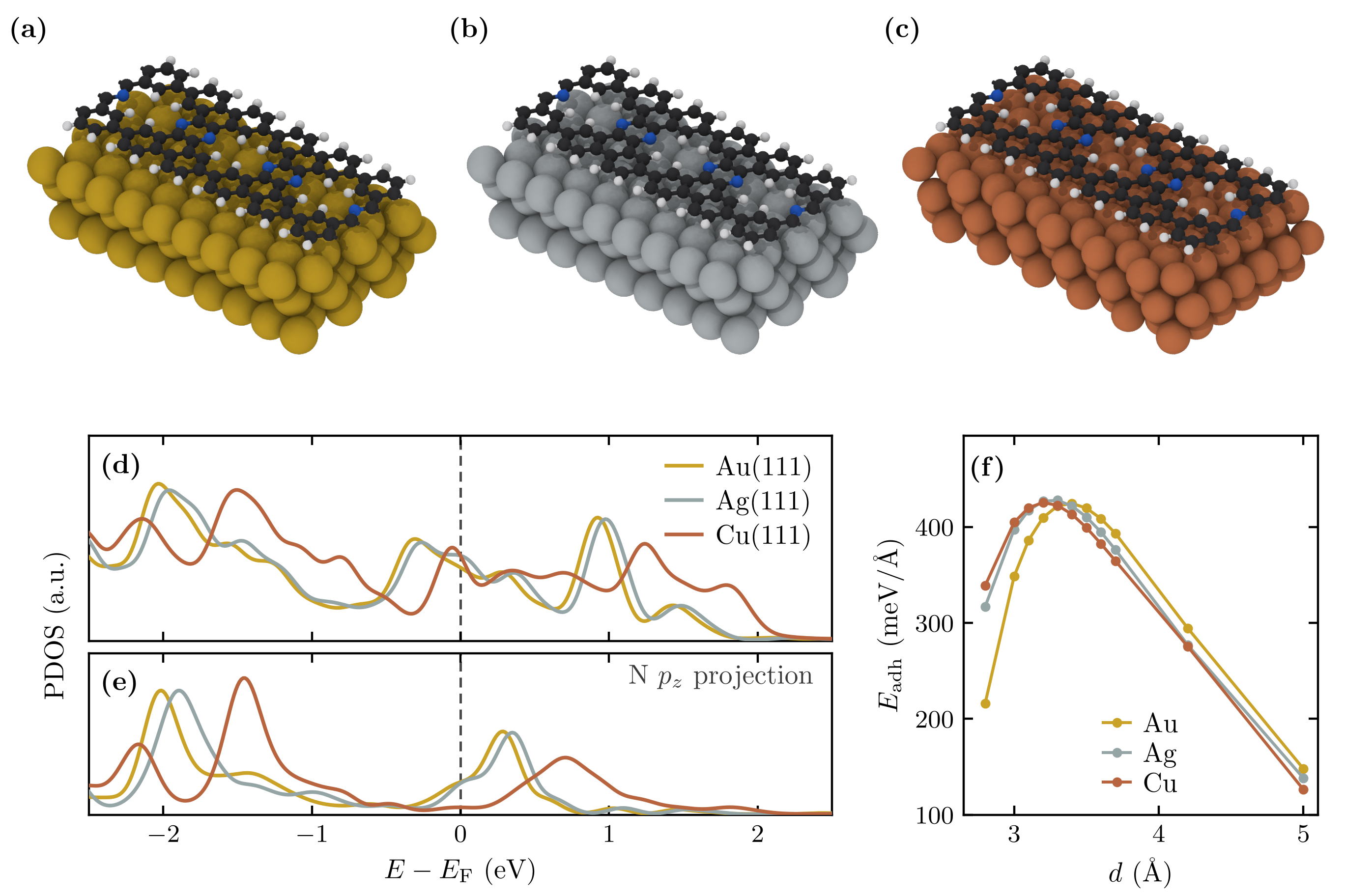}
  \caption{(a-c) The pyrazinic ribbon on Au(111), Ag(111) and Cu(111), each at its
equilibrium separation. (d) Total density of states projected on the ribbon and (e) on the nitrogen $p_z$ orbitals, relative to the Fermi level of each metal. (f) Adhesion energy against separation.}
  \label{fig:substrato}
\end{figure*}

Everything described so far refers to a free-standing ribbon. These ribbons are grown on a metal surface and characterized on it, so that the interface forms part of the system rather than an incidental detail of the measurement. Figure~\ref{fig:substrato}(a-c) shows the doped ribbon on the three noble metal (111) surfaces, \ref{fig:substrato}(d,e) the density of states at the interface and \ref{fig:substrato}(f) the adhesion curves.

The adhesion energy per unit length of ribbon in Figure~\ref{fig:substrato}(f),
\begin{equation}
E_\mathrm{adh}(d) = \frac{E_\mathrm{ribbon} + E_\mathrm{slab} - E_\mathrm{total}(d)}{a},
\label{eq:eadh}
\end{equation}
where $a$ is the length of the unit cell, peaks, by a parabolic fit through the three points nearest the maximum, at \SI{424.5}{\milli\electronvolt\per\angstrom} on Au(111), \SI{428.1}{\milli\electronvolt\per\angstrom} on Ag(111) and \SI{425.5}{\milli\electronvolt\per\angstrom} on Cu(111), or roughly \SI{88}{\milli\electronvolt} per atom, at separations of \num{3.4}, \num{3.3} and \SI{3.2}{\angstrom}. Both the magnitude and the distance identify physisorption. The curvature of these curves at the maximum corresponds to an out-of-plane restoring constant of \SIrange{0.14}{0.19}{\electronvolt\per\angstrom\squared} per atom, the stiffness with which the surface resists out-of-plane displacement of the ribbon throughout growth and characterization. The density of states projected on the ribbon in \ref{fig:substrato}(d) is nearly the same on gold and silver and departs from both on copper, so the energetic equivalence of the three supports does not extend to the alignment of the ribbon states with the metal. The scatter among the three metals is \SI{0.8}{\percent}, and the equilibrium separations decrease by \SI{0.1}{\angstrom} from gold to silver to copper, a spread of the order of the sampling itself. Benzene on the same three surfaces binds in the same regime and with a similar insensitivity, whereas the reactive transition metals bind chemically \citep{Liu2015,Vanin2010,Olsen2013}. Because dispersion does not distinguish among noble metals, no substrate-specific bonding requirement has to be met, which is consistent with the tolerance observed in synthesis, where the same ribbon forms on Cu(111), Ag(100), Ag(110) and Ag(111) \citep{Ceccatto2026}. The separations obtained here are longer by about \SI{0.2}{\angstrom} than the value reported for Ag(111) \citep{Pawlak2020}, which is the offset documented for this class of van der Waals functional \citep{Klimes2010}.

The nitrogen site, by contrast, does distinguish among the substrates. The nitrogen $p_z$ projection of Figure~\ref{fig:substrato}(e), integrated within half an electronvolt of the Fermi level and normalized identically for the three metals, gives \num{15.5} on Au and \num{14.8} on Ag against only \num{5.1} on Cu. On copper the ribbon states also shift upward by \SIrange{0.3}{0.5}{\electronvolt}, reversing which side of the Fermi level carries the larger weight. Copper has its $d$ band closest to the Fermi level of the three, and zigzag ribbons on the same three surfaces have been shown to hybridize more strongly with copper than with silver or gold \citep{Li2013}. The pyrazinic site therefore reports on the support even where the adhesion energy does not discriminate, and spectroscopy resolved on the nitrogen should distinguish copper from silver in samples that already exist.

\begin{figure*}[pos = !t]
  \centering
  \includegraphics[width=\linewidth]{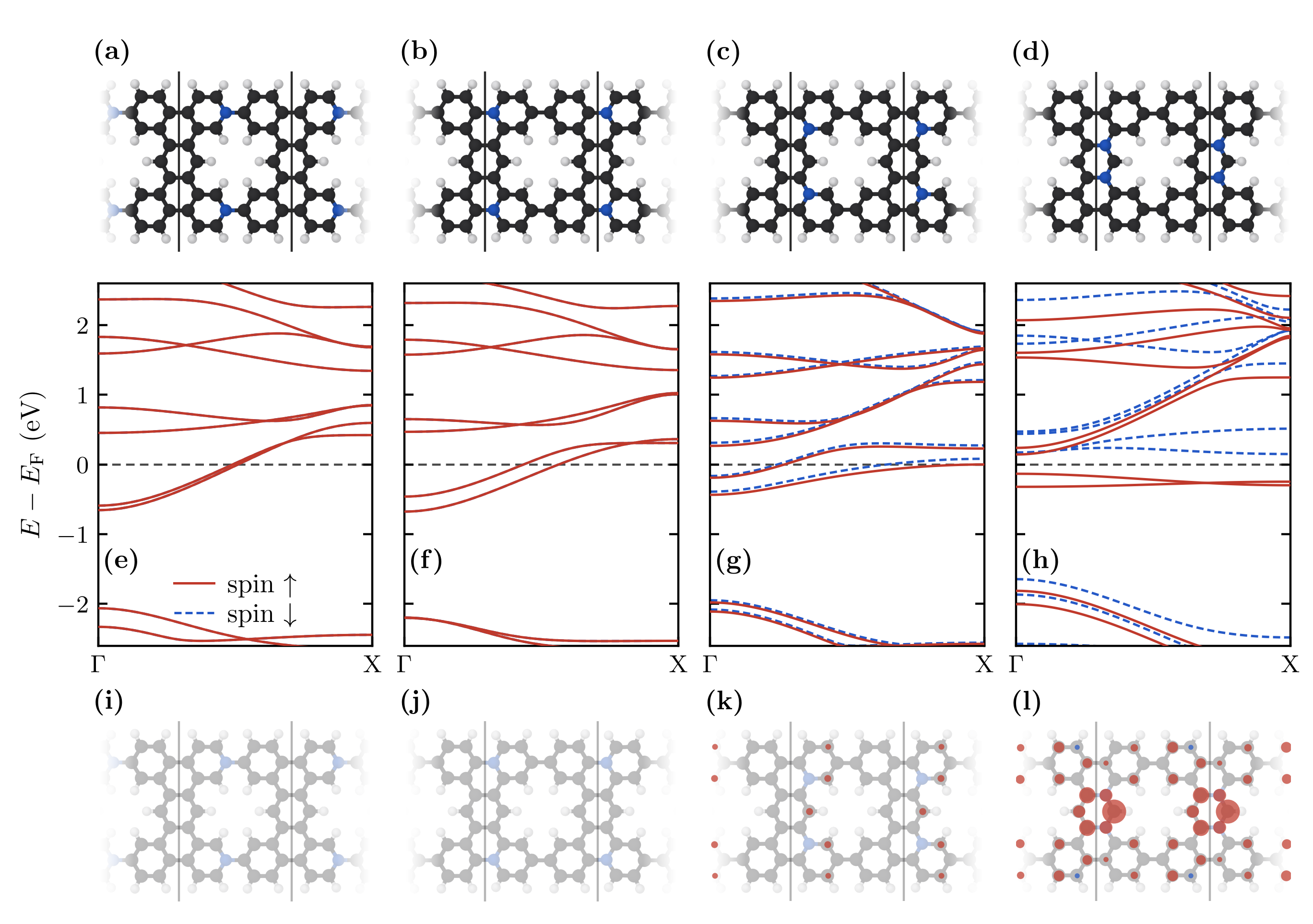}
  \caption{Four graphitic configurations differing only in the position of the
two nitrogen atoms across the ribbon width. (a-d) Structures, (e-h) spin-resolved band structures and (i-l) atom-resolved local moments.}
  \label{fig:grafiticas}
\end{figure*}

The synthesis delivers the pyrazinic site, and the behavior of the same ribbon with nitrogen at graphitic sites is accessible only to calculation. Figure~\ref{fig:grafiticas}(a-d) addresses that extension with four configurations that differ only in where the two nitrogen atoms sit across the ribbon width, together with their spin-resolved bands in \ref{fig:grafiticas}(e-h) and their atom-resolved moments in \ref{fig:grafiticas}(i-l).

Graphitic nitrogen is three-coordinated, and its C-N bonds measure between \num{1.39} and \SI{1.43}{\angstrom}, nearly indistinguishable from the aromatic carbon-carbon bonds of the same lattice. Chemically it behaves almost as a carbon atom, except that with two such atoms per cell the two donated electrons raise the Fermi level into the conduction manifold. In the three configurations where the nitrogen atoms sit away from the pore the Fermi level lies inside the conduction manifold and the bands in Figure~\ref{fig:grafiticas}(e-g) close to within \SI{20}{\milli\electronvolt}, at the limit of what the calculation resolves. This is the behavior established for graphitic nitrogen in two-dimensional graphene and in core-doped ribbons \citep{Zhao2011,Wen2023,Cervantes2008}, in contrast with the behavior of the pyrazinic site. Since the experiment observes a large gap and no states at the Fermi level, the calculation supports the pyrazinic assignment made from the structure.

Energy orders the four configurations by how far the nitrogen sits from the pore. Taking the most stable as reference, moving the pair one row inward costs \SI{467}{\milli\electronvolt} and moving it to the pore rim costs \SI{615}{\milli\electronvolt}, against only \SI{73}{\milli\electronvolt} between two inequivalent sites of the same row. Graphitic nitrogen prefers a fully coordinated aromatic environment, and the pore edge is neither.

Magnetism appears only when the two nitrogen atoms approach each other. The total moment is zero in the two outer configurations, \num{0.26} Bohr magnetons when the separation falls to \SI{4.99}{\angstrom} and \num{2.00} when it falls to \SI{2.46}{\angstrom}. In that last configuration the ferromagnetic solution lies \SI{35}{\milli\electronvolt} below the antiferromagnetic one and \SI{184}{\milli\electronvolt} below the non-magnetic one, so the moment itself is robust at room temperature while the alignment between the two sites is separated from the antiferromagnetic arrangement by little more than the thermal energy. The mechanism is the one established for defect-induced magnetism in carbon \citep{Yazyev2007,Yazyev2010,Nair2012}. The donated electrons occupy nearly degenerate states. As the nitrogen atoms approach, the density concentrates on a narrow set of states at the pore rim, the density of states at the Fermi level rises and the Stoner criterion is satisfied. The dependence on the distance between dopants follows what has been measured \citep{Tison2015} and calculated \citep{Yutomo2021} for nitrogen pairs in graphene.

The moment does not reside on the nitrogen. The atom-resolved moments in Figure~\ref{fig:grafiticas}(i-l) place \num{1.74} of the two Bohr magnetons on the carbon frame and \num{0.27} on the two nitrogen atoms together, the small remainder being a negative residue on the hydrogen. The largest individual moment, \num{0.46}, sits on a carbon atom of the central row at the rim of the pore. The nitrogen supplies the electrons and the carbon lattice hosts the moment, a division of roles also reported when graphitic nitrogen triggers ferromagnetism in graphene \citep{Blonski2017,Blackwell2021}. The band structure of that configuration, in Figure~\ref{fig:grafiticas}(h), is strongly spin asymmetric, with a gap of \SI{0.28}{\electronvolt} in one channel and \SI{1.80}{\electronvolt} in the other, so that states at the lower band edge belong to a single spin channel.

The cost of \SI{615}{\milli\electronvolt} relative to the outer configuration places this structure beyond the reach of a thermodynamically controlled synthesis, which is why the existing route delivers the pyrazinic ribbon. The configuration remains a well-defined target for a precursor designed to place the nitrogen at the pore rim, in the manner in which the pyrazine ring of the present precursor fixes the two-coordinated site.

\section{Conclusions}\label{sec:conclusoes}

We have characterized the synthesized pyrazinic nitrogen-doped porous armchair graphene nanoribbon across structure, dynamical and thermal stability, mechanical response, thermal transport, electronic structure, optical response including electron-hole interaction, and adsorption on noble metal surfaces, in each case alongside the pristine and the undoped porous ribbon.

The pore is the structural perturbation, and it is opened exactly where the frontier states of the pristine ribbon have their largest amplitude. It costs \SI{41}{\percent} of the stiffness, removes a factor of \num{3.7} from the thermal conductivity, mostly by creating internal surface that scatters phonons, and opens the gap by more than one electronvolt. The nitrogen, by contrast, leaves the mechanics unchanged to within \SI{1}{\percent}, removes a further \SI{23}{\percent} of the conductivity through the change it makes to the phonon spectrum, and acts on the conduction band alone, where it concentrates \SI{29}{\percent} of the state on two sites while leaving the valence band on the carbon frame.

Together the two shift the absorption edge from the infrared to the red, at \SI{646}{\nano\meter}, and bind the first exciton by \SI{414}{\milli\electronvolt}, the largest of the three ribbons, with the partial charge-transfer character that the donor-acceptor asymmetry implies. The hybrid functional places the gap at \SI{2.34}{\electronvolt}, within the window measured by STS, and predicts that an optical measurement should find a value lower by the exciton binding energy.

The effective phonon mean free path of the doped ribbon, \SI{9.9}{\nano\meter}, coincides with the lengths the synthesis currently reaches, so the intrinsic conductivity is accessible only by extrapolating a length series and not by measuring a single ribbon. The ribbon also physisorbs on Au(111), Ag(111) and Cu(111) with energies that differ by less than one percent. This accounts for the substrate tolerance observed in synthesis and supports transfer of the system to other supports. The nitrogen site nevertheless remains sensitive enough to distinguish copper from the other two.

Finally, graphitic rather than pyrazinic nitrogen would render the same ribbon metallic. Bringing the two nitrogen atoms to the pore edge would instead produce a ferromagnetic semiconductor with two Bohr magnetons per cell, a strongly spin-asymmetric gap, and the ferromagnetic alignment favored over the antiferromagnetic one by \SI{35}{\milli\electronvolt}. In that structure the moment resides on the carbon lattice rather than on the heteroatom. That configuration lies \SI{615}{\milli\electronvolt} above the one that thermodynamics prefers, which defines it as a target for precursor design rather than a description of what the current route produces.

\section*{CRediT authorship contribution statement}
\noindent \textbf{Cicera M. V. de Ara\'ujo:} Methodology, Investigation, Data curation, Writing -- original draft.
\textbf{Isaac de M. F\'elix:} Methodology, Software, Validation, Formal analysis, Data curation, Writing -- review \& editing.
\textbf{Willian F. Radel:} Methodology, Software, Validation, Data curation, Writing -- original draft.
\textbf{Raphael B. de Oliveira:} Methodology, Software, Validation, Investigation, Writing -- original draft.
\textbf{Guilherme da S. L. Fabris:} Methodology, Software, Validation, Formal analysis, Investigation, Writing -- review \& editing.
\textbf{Douglas S. Galv\~ao:} Validation, Formal analysis, Investigation, Resources, Supervision, Funding acquisition, Writing -- review \& editing.
\textbf{Marcelo L. Pereira Junior:} Conceptualization, Methodology, Software, Formal analysis, Investigation, Resources, Data curation, Writing -- review \& editing, Visualization, Supervision, Project administration, Funding acquisition.

\section*{Declaration of competing interest}
The authors declare that they have no known competing financial interests or personal relationships that could have appeared to influence the work reported in this paper.

\section*{Data availability}
Data will be made available on request.

\section*{Acknowledgements}
\noindent C.M.V.A. thanks CAPES process number 88887.702709/ 2022-00. W.F.R. thanks CAPES process number 88887. 840299/2023-00. G.S.L.F.\ acknowledges the S\~ao Paulo Research Foundation (FAPESP) fellowship (process number 2024/03413-9). R.B.O. thanks CNPq process number \#151043/2024-8. D.S.G.\ acknowledges the Center for Computing in Engineering and Sciences at Unicamp for financial support through the FAPESP CEPID Grant (process number 2013/08293-7) and support from INEO/CNPq and FAPESP (grant 2025/27044-5).  
M.L.P.J. acknowledges financial support from FAPDF (grant 00193-00001807/\allowbreak 2023-16), CNPq (grants 444921/\allowbreak 2024-9 and 308222/\allowbreak 2025-3), and CAPES (grant 88887.\allowbreak 005164/\allowbreak 2024-00).

\bibliographystyle{elsarticle-num}
\bibliography{refs}

@article{Alves2025, author={Alves, Rodrigo A. F. and Lima, Kleuton A. L. and da Silva, Daniel A. and Mendonça, Fábio L. L. and Ribeiro Junior, Luiz A. and Pereira Junior, Marcelo L.}, title={Computational Design of 2D Nanoporous Graphene via Carbon-Bridged Lateral Heterojunctions in Armchair Graphene Nanoribbons}, journal={ACS Omega}, volume={10}, pages={17159--17169}, year={2025}, doi={10.1021/acsomega.4c07524}}

@article{Balandin2008, author={Balandin, Alexander A. and Ghosh, Suchismita and Bao, Wenzhong and Calizo, Irene and Teweldebrhan, Desalegne and Miao, Feng and Lau, Chun Ning}, title={Superior Thermal Conductivity of Single-Layer Graphene}, journal={Nano Letters}, volume={8}, pages={902--907}, year={2008}, doi={10.1021/nl0731872}}

@article{Balandin2011, author={Balandin, Alexander A.}, title={Thermal properties of graphene and nanostructured carbon materials}, journal={Nature Materials}, volume={10}, pages={569--581}, year={2011}, doi={10.1038/nmat3064}}

@article{Bassi2024, author={Bassi, Nicol{\`o} and Xu, Xiushang and Xiang, Feifei and Krane, Nils and Pignedoli, Carlo A. and Narita, Akimitsu and Fasel, Roman and Ruffieux, Pascal}, title={Preferential graphitic-nitrogen formation in pyridine-extended graphene nanoribbons}, journal={Communications Chemistry}, volume={7}, pages={274}, year={2024}, doi={10.1038/s42004-024-01344-7}}

@article{Bessa2025, author={Bessa, Mizraim and de Medeiros Dantas, D{\^e}nis G. and Gomes, Djardiel da Silva and Pereira Junior, Marcelo L. and Azevedo, S{\'e}rgio and Machado, Leonardo D.}, title={Mechanical strength and strain-induced optical shifts in monolayer azugraphene}, journal={Computational Materials Science}, volume={258}, pages={114087}, year={2025}, doi={10.1016/j.commatsci.2025.114087}}

@article{Bieri2009, author={Bieri, Marco and Treier, Matthias and Cai, Jinming and Aït-Mansour, Kamel and Ruffieux, Pascal and Gröning, Oliver and Gröning, Pierangelo and Kastler, Marcel and Rieger, Ralph and Feng, Xinliang and Müllen, Klaus and Fasel, Roman}, title={Porous graphenes: two-dimensional polymer synthesis with atomic precision}, journal={Chemical Communications}, volume={45}, pages={6919--6921}, year={2009}, doi={10.1039/b915190g}}

@article{Blackwell2021, author={Blackwell, Raymond E. and Zhao, Fangzhou and Brooks, Erin and Zhu, Junmian and Piskun, Ilya and Wang, Shenkai and Delgado, Aidan and Lee, Yea-Lee and Louie, Steven G. and Fischer, Felix R.}, title={Spin splitting of dopant edge state in magnetic zigzag graphene nanoribbons}, journal={Nature}, volume={600}, pages={647--652}, year={2021}, doi={10.1038/s41586-021-04201-y}}

@article{Blonski2017, author={B{\l}o{\'n}ski, Piotr and Tu{\v c}ek, Ji{\v r}{\'i} and Sofer, Zden{\v e}k and Maz{\'a}nek, Vlastimil and Petr, Martin and Pumera, Martin and Otyepka, Michal and Zbo{\v r}il, Radek}, title={Doping with Graphitic Nitrogen Triggers Ferromagnetism in Graphene}, journal={Journal of the American Chemical Society}, volume={139}, pages={3171--3180}, year={2017}, doi={10.1021/jacs.6b12934}}

@article{Cai2010, author={Cai, Jinming and Ruffieux, Pascal and Jaafar, Rached and Bieri, Marco and Braun, Thomas and Blankenburg, Stephan and Muoth, Matthias and Seitsonen, Ari P. and Saleh, Moussa and Feng, Xinliang and Müllen, Klaus and Fasel, Roman}, title={Atomically precise bottom-up fabrication of graphene nanoribbons}, journal={Nature}, volume={466}, pages={470--473}, year={2010}, doi={10.1038/nature09211}}

@article{Cai2014, author={Cai, Jinming and Pignedoli, Carlo A. and Talirz, Leopold and Ruffieux, Pascal and Söde, Hajo and Liang, Liangbo and Meunier, Vincent and Berger, Reinhard and Li, Rongjin and Feng, Xinliang and Müllen, Klaus and Fasel, Roman}, title={Graphene nanoribbon heterojunctions}, journal={Nature Nanotechnology}, volume={9}, pages={896--900}, year={2014}, doi={10.1038/nnano.2014.184}}

@article{CastroNeto2009, author={Castro Neto, A. H. and Guinea, F. and Peres, N. M. R. and Novoselov, K. S. and Geim, A. K.}, title={The electronic properties of graphene}, journal={Reviews of Modern Physics}, volume={81}, pages={109--162}, year={2009}, doi={10.1103/RevModPhys.81.109}}

@article{Ceccatto2026, author={Ceccatto, Alisson and Herrera-Reinoza, Nataly and Silva, Marcela C. R. and Pérez Paz, Alejandro and Carreño Díaz, Vanessa and Pilli, Ronaldo Aloise and de Siervo, Abner}, title={On-surface synthesis of porous nitrogen-doped graphene nanoribbons: Role of substrate orientation}, journal={Carbon}, volume={259}, pages={121895}, year={2026}, doi={10.1016/j.carbon.2026.121895}}

@article{Celebi2014, author={Celebi, Kemal and Buchheim, Jakob and Wyss, Roman M. and Droudian, Amirhossein and Gasser, Patrick and Shorubalko, Ivan and Kye, Jeong-Il and Lee, Changho and Park, Hyung Gyu}, title={Ultimate Permeation Across Atomically Thin Porous Graphene}, journal={Science}, volume={344}, pages={289--292}, year={2014}, doi={10.1126/science.1249097}}

@article{Cervantes2008, author={Cervantes-Sodi, F. and Csányi, G. and Piscanec, S. and Ferrari, A. C.}, title={Edge-functionalized and substitutionally doped graphene nanoribbons: Electronic and spin properties}, journal={Physical Review B}, volume={77}, pages={165427}, year={2008}, doi={10.1103/PhysRevB.77.165427}}

@article{Chen2012, author={Chen, Shanshan and Wu, Qingzhi and Mishra, Columbia and Kang, Junyong and Zhang, Hengji and Cho, Kyeongjae and Cai, Weiwei and Balandin, Alexander A. and Ruoff, Rodney S.}, title={Thermal conductivity of isotopically modified graphene}, journal={Nature Materials}, volume={11}, pages={203--207}, year={2012}, doi={10.1038/nmat3207}}

@article{daCunha2021, author={da Cunha, Wiliam F. and Pereira Júnior, Marcelo L. and Giozza, William F. and de Sousa Junior, Rafael T. and Ribeiro Júnior, Luiz A. and e Silva, Geraldo M.}, title={Polaron transport in porous graphene nanoribbons}, journal={Computational Materials Science}, volume={194}, pages={110423}, year={2021}, doi={10.1016/j.commatsci.2021.110423}}

@article{Denk2014, author={Denk, Richard and Hohage, Michael and Zeppenfeld, Peter and Cai, Jinming and Pignedoli, Carlo A. and S{\"o}de, Hajo and Fasel, Roman and Feng, Xinliang and M{\"u}llen, Klaus and Wang, Shudong and Prezzi, Deborah and Ferretti, Andrea and Ruini, Alice and Molinari, Elisa and Ruffieux, Pascal}, title={Exciton-dominated optical response of ultra-narrow graphene nanoribbons}, journal={Nature Communications}, volume={5}, pages={4253}, year={2014}, doi={10.1038/ncomms5253}}

@article{DeSousa2026, author={De Sousa, José M. and Pereira Junior, Marcelo L. and Galvão, Douglas S. and Ribeiro Junior, Luiz A. and Fonseca, Alexandre F.}, title={Chirality and width effects on elastic and fracture properties of nanoribbons with coronene edges}, journal={Scientific Reports}, volume={16}, pages={24087}, year={2026}, doi={10.1038/s41598-026-55137-0}}

@article{Dias2023, author={Dias, Alexandre C. and Silveira, Julian F. R. V. and Qu, Fanyao}, title={WanTiBEXOS: A Wannier based Tight Binding code for electronic band structure, excitonic and optoelectronic properties of solids}, journal={Computer Physics Communications}, volume={285}, pages={108636}, year={2023}, doi={10.1016/j.cpc.2022.108636}}

@article{Dion2004, author={Dion, M. and Rydberg, H. and Schröder, E. and Langreth, D. C. and Lundqvist, B. I.}, title={Van der Waals Density Functional for General Geometries}, journal={Physical Review Letters}, volume={92}, pages={246401}, year={2004}, doi={10.1103/PhysRevLett.92.246401}}

@article{Evans2010, author={Evans, William J. and Hu, Lin and Keblinski, Pawel}, title={Thermal conductivity of graphene ribbons from equilibrium molecular dynamics: Effect of ribbon width, edge roughness, and hydrogen termination}, journal={Applied Physics Letters}, volume={96}, pages={203112}, year={2010}, doi={10.1063/1.3435465}}

@article{Fan2015, author={Fan, Qitang and Gottfried, J. Michael and Zhu, Junfa}, title={Surface-Catalyzed C--C Covalent Coupling Strategies toward the Synthesis of Low-Dimensional Carbon-Based Nanostructures}, journal={Accounts of Chemical Research}, volume={48}, pages={2484--2494}, year={2015}, doi={10.1021/acs.accounts.5b00168}}

@article{Fan2024, author={Fan, Qitang and Ruan, Zilin and Werner, Simon and Naumann, Tim and Bolat, Rustem and Martinez-Castro, Jose and Koehler, Tabea and Vollgraff, Tobias and Hieringer, Wolfgang and Mandalia, Raviraj and Neiß, Christian and Görling, Andreas and Tautz, F. Stefan and Sundermeyer, Jörg and Gottfried, J. Michael}, title={Bottom-up Synthesis and Characterization of Porous 12-Atom-Wide Armchair Graphene Nanoribbons}, journal={Nano Letters}, volume={24}, pages={10718--10723}, year={2024}, doi={10.1021/acs.nanolett.4c01106}}

@article{Feng2016, author={Feng, Tianli and Ruan, Xiulin}, title={Ultra-low thermal conductivity in graphene nanomesh}, journal={Carbon}, volume={101}, pages={107--113}, year={2016}, doi={10.1016/j.carbon.2016.01.082}}

@article{Garcia2020, author={Garc{\'i}a, Alberto and Papior, Nick and Akhtar, Arsalan and Artacho, Emilio and Blum, Volker and Bosoni, Emanuele and Brandimarte, Pedro and Brandbyge, Mads and Cerd{\'a}, J. I. and Corsetti, Fabiano and Cuadrado, Ram{\'o}n and Dikan, Vladimir and Ferrer, Jaime and Gale, Julian and Garc{\'i}a-Fern{\'a}ndez, Pablo and Garc{\'i}a-Su{\'a}rez, V. M. and Garc{\'i}a, Sandra and Huhs, Georg and Illera, Sergio and Koryt{\'a}r, Richard and Koval, Peter and Lebedeva, Irina and Lin, Lin and L{\'o}pez-Tarifa, Pablo and Mayo, Sara G. and Mohr, Stephan and Ordej{\'o}n, Pablo and Postnikov, Andrei and Pouillon, Yann and Pruneda, Miguel and Robles, Roberto and S{\'a}nchez-Portal, Daniel and Soler, Jos{\'e} M. and Ullah, Rafi and Yu, Victor Wen-zhe and Junquera, Javier}, title={Siesta: Recent developments and applications}, journal={The Journal of Chemical Physics}, volume={152}, pages={204108}, year={2020}, doi={10.1063/5.0005077}}

@article{Geim2007, author={Geim, A. K. and Novoselov, K. S.}, title={The rise of graphene}, journal={Nature Materials}, volume={6}, pages={183--191}, year={2007}, doi={10.1038/nmat1849}}

@article{Gomes2025, author={Gomes, Djardiel S. and Felix, Isaac M. and Radel, Willian F. and Dias, Alexandre C. and Ribeiro Junior, Luiz A. and Pereira Junior, Marcelo L.}, title={Computational Characterization of the Recently Synthesized Pristine and Porous 12-Atom-Wide Armchair Graphene Nanoribbon}, journal={Nano Letters}, volume={25}, pages={8596--8603}, year={2025}, doi={10.1021/acs.nanolett.5c01319}}

@article{Gomes2026, author={da Silva Gomes, Djardiel and Felix, Isaac Macedo and Lage, Lucas Lopes and Galvão, Douglas Soares and Latgé, Andrea and Pereira Junior, Marcelo Lopes}, title={Topology as a Design Variable for Multiproperty Engineering in Synthesized 4-5-6-8 Carbon Nanoribbons}, journal={ACS Materials Au}, year={2026}, doi={10.1021/acsmaterialsau.6c00131}}

@article{Grill2007, author={Grill, Leonhard and Dyer, Matthew and Lafferentz, Leif and Persson, Mats and Peters, Maike V. and Hecht, Stefan}, title={Nano-architectures by covalent assembly of molecular building blocks}, journal={Nature Nanotechnology}, volume={2}, pages={687--691}, year={2007}, doi={10.1038/nnano.2007.346}}

@article{Haskins2011, author={Haskins, Justin and K{\i}nac{\i}, Alper and Sevik, Cem and Sevin{\c c}li, H{\^a}ldun and Cuniberti, Gianaurelio and {\c C}a{\u g}{\i}n, Tahir}, title={Control of Thermal and Electronic Transport in Defect-Engineered Graphene Nanoribbons}, journal={ACS Nano}, volume={5}, pages={3779--3787}, year={2011}, doi={10.1021/nn200114p}}

@article{Heyd2003, author={Heyd, Jochen and Scuseria, Gustavo E. and Ernzerhof, Matthias}, title={Hybrid functionals based on a screened Coulomb potential}, journal={The Journal of Chemical Physics}, volume={118}, pages={8207--8215}, year={2003}, doi={10.1063/1.1564060}}

@article{Heyd2006erratum, author={Heyd, Jochen and Scuseria, Gustavo E. and Ernzerhof, Matthias}, title={Erratum: ``Hybrid functionals based on a screened Coulomb potential'' [J. Chem. Phys. 118, 8207 (2003)]}, journal={The Journal of Chemical Physics}, volume={124}, pages={219906}, year={2006}, doi={10.1063/1.2204597}}

@article{Hu2009, author={Hu, Jiuning and Ruan, Xiulin and Chen, Yong P.}, title={Thermal Conductivity and Thermal Rectification in Graphene Nanoribbons: A Molecular Dynamics Study}, journal={Nano Letters}, volume={9}, pages={2730--2735}, year={2009}, doi={10.1021/nl901231s}}

@article{Jiang2009, author={Jiang, De-en and Cooper, Valentino R. and Dai, Sheng}, title={Porous Graphene as the Ultimate Membrane for Gas Separation}, journal={Nano Letters}, volume={9}, pages={4019--4024}, year={2009}, doi={10.1021/nl9021946}}

@article{Joucken2015, author={Joucken, Fr{\'e}d{\'e}ric and Tison, Yann and Le F{\`e}vre, Patrick and Tejeda, Antonio and Taleb-Ibrahimi, Amina and Conrad, Edward and Repain, Vincent and Chacon, Cyril and Bellec, Amandine and Girard, Yann and Rousset, Sylvie and Ghijsen, Jacques and Sporken, Robert and Amara, Hakim and Ducastelle, Fran{\c c}ois and Lagoute, J{\'e}r{\^o}me}, title={Charge transfer and electronic doping in nitrogen-doped graphene}, journal={Scientific Reports}, volume={5}, pages={14564}, year={2015}, doi={10.1038/srep14564}}

@article{Junquera2001, author={Junquera, Javier and Paz, {\'O}scar and S{\'a}nchez-Portal, Daniel and Artacho, Emilio}, title={Numerical atomic orbitals for linear-scaling calculations}, journal={Physical Review B}, volume={64}, pages={235111}, year={2001}, doi={10.1103/PhysRevB.64.235111}}

@article{Kharche2016, author={Kharche, Neerav and Meunier, Vincent}, title={Width and Crystal Orientation Dependent Band Gap Renormalization in Substrate-Supported Graphene Nanoribbons}, journal={The Journal of Physical Chemistry Letters}, volume={7}, pages={1526--1533}, year={2016}, doi={10.1021/acs.jpclett.6b00422}}

@article{Klimes2010, author={Klime{\v s}, Ji{\v r}{\'i} and Bowler, David R. and Michaelides, Angelos}, title={Chemical accuracy for the van der Waals density functional}, journal={Journal of Physics: Condensed Matter}, volume={22}, pages={022201}, year={2010}, doi={10.1088/0953-8984/22/2/022201}}

@article{Knupfer2003, author={Knupfer, M.}, title={Exciton binding energies in organic semiconductors}, journal={Applied Physics A}, volume={77}, pages={623--626}, year={2003}, doi={10.1007/s00339-003-2182-9}}

@article{Kowalik2019, author={Kowalik, Malgorzata and Ashraf, Chowdhury and Damirchi, Behzad and Akbarian, Dooman and Rajabpour, Siavash and van Duin, Adri C. T.}, title={Atomistic Scale Analysis of the Carbonization Process for C/H/O/N-Based Polymers with the ReaxFF Reactive Force Field}, journal={The Journal of Physical Chemistry B}, volume={123}, pages={5357--5367}, year={2019}, doi={10.1021/acs.jpcb.9b04298}}

@article{Krukau2006, author={Krukau, Aliaksandr V. and Vydrov, Oleg A. and Izmaylov, Artur F. and Scuseria, Gustavo E.}, title={Influence of the exchange screening parameter on the performance of screened hybrid functionals}, journal={The Journal of Chemical Physics}, volume={125}, pages={224106}, year={2006}, doi={10.1063/1.2404663}}

@article{Lafferentz2012, author={Lafferentz, L. and Eberhardt, V. and Dri, C. and Africh, C. and Comelli, G. and Esch, F. and Hecht, S. and Grill, L.}, title={Controlling on-surface polymerization by hierarchical and substrate-directed growth}, journal={Nature Chemistry}, volume={4}, pages={215--220}, year={2012}, doi={10.1038/nchem.1242}}

@article{Lage2026, author={Lage, Lucas Lopes and Félix, A. B. and Gomes, Djardiel S. and Pereira Junior, Marcelo Lopes and Latgé, Andrea}, title={Emergent hierarchy in localized states of organic quantum chains}, journal={Nanoscale}, volume={18}, pages={11391--11400}, year={2026}, doi={10.1039/d6nr00440g}}

@article{Lazar2019, author={Lazar, Petr and Mach, Radim and Otyepka, Michal}, title={Spectroscopic Fingerprints of Graphitic, Pyrrolic, Pyridinic, and Chemisorbed Nitrogen in N-Doped Graphene}, journal={The Journal of Physical Chemistry C}, volume={123}, pages={10695--10702}, year={2019}, doi={10.1021/acs.jpcc.9b02163}}

@article{Lee2008, author={Lee, Changgu and Wei, Xiaoding and Kysar, Jeffrey W. and Hone, James}, title={Measurement of the Elastic Properties and Intrinsic Strength of Monolayer Graphene}, journal={Science}, volume={321}, pages={385--388}, year={2008}, doi={10.1126/science.1157996}}

@article{Li2013, author={Li, Yan and Zhang, Wei and Morgenstern, Markus and Mazzarello, Riccardo}, title={Electronic and Magnetic Properties of Zigzag Graphene Nanoribbons on the (111) Surface of Cu, Ag, and Au}, journal={Physical Review Letters}, volume={110}, pages={216804}, year={2013}, doi={10.1103/PhysRevLett.110.216804}}

@article{Lima2025, author={Lima, Kleuton A. L. and Alves, Rodrigo A. F. and da Silva, Daniel A. and Mendon{\c c}a, F{\'a}bio L. L. and Pereira Junior, Marcelo L. and Ribeiro Junior, Luiz A.}, title={TH-graphyne: a new porous bidimensional carbon allotrope}, journal={Physical Chemistry Chemical Physics}, volume={27}, pages={8684--8691}, year={2025}, doi={10.1039/D4CP02923B}}

@article{Lin2015, author={Lin, Yung-Chang and Teng, Po-Yuan and Yeh, Chao-Hui and Koshino, Masanori and Chiu, Po-Wen and Suenaga, Kazu}, title={Structural and Chemical Dynamics of Pyridinic-Nitrogen Defects in Graphene}, journal={Nano Letters}, volume={15}, pages={7408--7413}, year={2015}, doi={10.1021/acs.nanolett.5b02831}}

@article{Liu2007, author={Liu, Fang and Ming, Pingbing and Li, Ju}, title={Ab initio calculation of ideal strength and phonon instability of graphene under tension}, journal={Physical Review B}, volume={76}, pages={064120}, year={2007}, doi={10.1103/PhysRevB.76.064120}}

@article{Liu2015, author={Liu, Wei and Maaß, Friedrich and Willenbockel, Martin and Bronner, Christopher and Schulze, Michael and Soubatch, Serguei and Tautz, F. Stefan and Tegeder, Petra and Tkatchenko, Alexandre}, title={Quantitative Prediction of Molecular Adsorption: Structure and Binding of Benzene on Coinage Metals}, journal={Physical Review Letters}, volume={115}, pages={036104}, year={2015}, doi={10.1103/PhysRevLett.115.036104}}

@article{Llinas2017, author={Llinas, Juan Pablo and Fairbrother, Andrew and Borin Barin, Gabriela and Shi, Wu and Lee, Kyunghoon and Wu, Shuang and Choi, Byung Yong and Braganza, Rohit and Lear, Jordan and Kau, Nicholas and Choi, Wonwoo and Chen, Chen and Pedramrazi, Zahra and Dumslaff, Tim and Narita, Akimitsu and Feng, Xinliang and M{\"u}llen, Klaus and Fischer, Felix and Zettl, Alex and Ruffieux, Pascal and Yablonovitch, Eli and Crommie, Michael and Fasel, Roman and Bokor, Jeffrey}, title={Short-channel field-effect transistors with 9-atom and 13-atom wide graphene nanoribbons}, journal={Nature Communications}, volume={8}, pages={633}, year={2017}, doi={10.1038/s41467-017-00734-x}}

@article{MerinoDiez2017, author={Merino-D{\'i}ez, N{\'e}stor and Garcia-Lekue, Aran and Carbonell-Sanrom{\`a}, Eduard and Li, Jingcheng and Corso, Martina and Colazzo, Luciano and Sedona, Francesco and S{\'a}nchez-Portal, Daniel and Pascual, Jose Ignacio and de Oteyza, Dimas G.}, title={Width-Dependent Band Gap in Armchair Graphene Nanoribbons Reveals Fermi Level Pinning on Au(111)}, journal={ACS Nano}, volume={11}, pages={11661--11668}, year={2017}, doi={10.1021/acsnano.7b06765}}

@article{Monkhorst1976, author={Monkhorst, Hendrik J. and Pack, James D.}, title={Special points for Brillouin-zone integrations}, journal={Physical Review B}, volume={13}, pages={5188--5192}, year={1976}, doi={10.1103/PhysRevB.13.5188}}

@article{Moreno2018, author={Moreno, César and Vilas-Varela, Manuel and Kretz, Bernhard and Garcia-Lekue, Aran and Costache, Marius V. and Paradinas, Markos and Panighel, Mirko and Ceballos, Gustavo and Valenzuela, Sergio O. and Peña, Diego and Mugarza, Aitor}, title={Bottom-up synthesis of multifunctional nanoporous graphene}, journal={Science}, volume={360}, pages={199--203}, year={2018}, doi={10.1126/science.aar2009}}

@article{MullerPlathe1997, author={Müller-Plathe, Florian}, title={A simple nonequilibrium molecular dynamics method for calculating the thermal conductivity}, journal={The Journal of Chemical Physics}, volume={106}, pages={6082--6085}, year={1997}, doi={10.1063/1.473271}}

@article{Nair2012, author={Nair, R. R. and Sepioni, M. and Tsai, I-Ling and Lehtinen, O. and Keinonen, J. and Krasheninnikov, A. V. and Thomson, T. and Geim, A. K. and Grigorieva, I. V.}, title={Spin-half paramagnetism in graphene induced by point defects}, journal={Nature Physics}, volume={8}, pages={199--202}, year={2012}, doi={10.1038/nphys2183}}

@article{Nakada1996, author={Nakada, Kyoko and Fujita, Mitsutaka and Dresselhaus, Gene and Dresselhaus, Mildred S.}, title={Edge state in graphene ribbons: Nanometer size effect and edge shape dependence}, journal={Physical Review B}, volume={54}, pages={17954--17961}, year={1996}, doi={10.1103/PhysRevB.54.17954}}

@article{Nose1984, author={Nos{\'e}, Shuichi}, title={A unified formulation of the constant temperature molecular dynamics methods}, journal={The Journal of Chemical Physics}, volume={81}, pages={511--519}, year={1984}, doi={10.1063/1.447334}}

@article{Novoselov2004, author={Novoselov, K. S. and Geim, A. K. and Morozov, S. V. and Jiang, D. and Zhang, Y. and Dubonos, S. V. and Grigorieva, I. V. and Firsov, A. A.}, title={Electric Field Effect in Atomically Thin Carbon Films}, journal={Science}, volume={306}, pages={666--669}, year={2004}, doi={10.1126/science.1102896}}

@article{Olsen2013, author={Olsen, Thomas and Thygesen, Kristian S.}, title={Random phase approximation applied to solids, molecules, and graphene-metal interfaces: From van der Waals to covalent bonding}, journal={Physical Review B}, volume={87}, pages={075111}, year={2013}, doi={10.1103/PhysRevB.87.075111}}

@article{Parlinski1997, author={Parlinski, K. and Li, Z. Q. and Kawazoe, Y.}, title={First-Principles Determination of the Soft Mode in Cubic ZrO2}, journal={Physical Review Letters}, volume={78}, pages={4063--4066}, year={1997}, doi={10.1103/PhysRevLett.78.4063}}

@article{Paupitz2026, author={Paupitz, R. and Fonseca, A. F. and Bessa, M. and Fabris, G. S. L. and da Cunha, W. F. and Machado, L. D. and Pereira Junior, M. L. and Ribeiro Junior, L. A. and Galv{\~a}o, D. S.}, title={A concise review of recently synthesized 2D carbon allotropes: Amorphous carbon, graphynes, biphenylene and fullerene networks}, journal={Carbon}, volume={252}, pages={121320}, year={2026}, doi={10.1016/j.carbon.2026.121320}}

@article{Pawlak2020, author={Pawlak, Rémy and Liu, Xunshan and Ninova, Silviya and D'Astolfo, Philipp and Drechsel, Carl and Sangtarash, Sara and Häner, Robert and Decurtins, Silvio and Sadeghi, Hatef and Lambert, Colin J. and Aschauer, Ulrich and Liu, Shi-Xia and Meyer, Ernst}, title={Bottom-up Synthesis of Nitrogen-Doped Porous Graphene Nanoribbons}, journal={Journal of the American Chemical Society}, volume={142}, pages={12568--12573}, year={2020}, doi={10.1021/jacs.0c03946}}

@article{Pedersen2008, author={Pedersen, Thomas G. and Flindt, Christian and Pedersen, Jesper and Mortensen, Niels Asger and Jauho, Antti-Pekka and Pedersen, Kjeld}, title={Graphene Antidot Lattices: Designed Defects and Spin Qubits}, journal={Physical Review Letters}, volume={100}, pages={136804}, year={2008}, doi={10.1103/PhysRevLett.100.136804}}

@article{Perdew1996, author={Perdew, John P. and Burke, Kieron and Ernzerhof, Matthias}, title={Generalized Gradient Approximation Made Simple}, journal={Physical Review Letters}, volume={77}, pages={3865--3868}, year={1996}, doi={10.1103/PhysRevLett.77.3865}}

@article{Pereira2020NPG, author={Pereira Junior, Marcelo L. and Ribeiro Junior, Luiz A.}, title={Thermomechanical insight into the stability of nanoporous graphene membranes}, journal={FlatChem}, volume={24}, pages={100196}, year={2020}, doi={10.1016/j.flatc.2020.100196}}

@article{Pereira2022biphenylene, author={Pereira Junior, Marcelo L. and da Cunha, Wiliam F. and de Sousa Junior, Rafael T. and Amvame Nze, Georges D. and Galv{\~a}o, Douglas S. and Ribeiro Junior, Luiz A.}, title={On the mechanical properties and fracture patterns of the nonbenzenoid carbon allotrope (biphenylene network): a reactive molecular dynamics study}, journal={Nanoscale}, volume={14}, pages={3200--3211}, year={2022}, doi={10.1039/D1NR07959J}}

@article{PereiraJr2020chevron, author={Pereira Junior, Marcelo L. and da Cunha, Wiliam F. and de Sousa Junior, Rafael T. and Giozza, William F. and e Silva, Geraldo M. and Ribeiro Junior, Luiz A.}, title={Charge Transport Mechanism in Chevron-Graphene Nanoribbons}, journal={The Journal of Physical Chemistry C}, volume={124}, pages={22392--22398}, year={2020}, doi={10.1021/acs.jpcc.0c06625}}

@article{PereiraJr2020coronene, author={Pereira Junior, Marcelo L. and Enders Neto, Bernhard G. and Giozza, William F. and de Sousa Junior, Rafael T. and e Silva, Geraldo M. and Ribeiro Junior, Luiz A.}, title={Transport of quasiparticles in coronene-based graphene nanoribbons}, journal={Journal of Materials Chemistry C}, volume={8}, pages={12100--12107}, year={2020}, doi={10.1039/D0TC01319F}}

@article{PereiraJr2020necklace, author={Pereira Junior, Marcelo L. and e Silva, Geraldo M. and Ribeiro Junior, Luiz A.}, title={Bosonic Charge Carriers in Necklace-like Graphene Nanoribbons}, journal={The Journal of Physical Chemistry Letters}, volume={11}, pages={5538--5543}, year={2020}, doi={10.1021/acs.jpclett.0c01489}}

@article{PereiraJr2021topo, author={Pereira Junior, Marcelo L. and de Oliveira Neto, Pedro H. and da Silva Filho, Dem{\'e}trio A. and de Sousa, Leonardo E. and e Silva, Geraldo M. and Ribeiro Junior, Luiz A.}, title={Charge localization and hopping in a topologically engineered graphene nanoribbon}, journal={Scientific Reports}, volume={11}, pages={5142}, year={2021}, doi={10.1038/s41598-021-84626-7}}

@article{Plimpton1995, author={Plimpton, Steve}, title={Fast Parallel Algorithms for Short-Range Molecular Dynamics}, journal={Journal of Computational Physics}, volume={117}, pages={1--19}, year={1995}, doi={10.1006/jcph.1995.1039}}

@article{Prezzi2008, author={Prezzi, Deborah and Varsano, Daniele and Ruini, Alice and Marini, Andrea and Molinari, Elisa}, title={Optical properties of graphene nanoribbons: The role of many-body effects}, journal={Physical Review B}, volume={77}, pages={041404}, year={2008}, doi={10.1103/PhysRevB.77.041404}}

@article{Qin2015, author={Qin, Xinming and Shang, Honghui and Xiang, Hongjun and Li, Zhenyu and Yang, Jinlong}, title={HONPAS: A linear scaling open-source solution for large system simulations}, journal={International Journal of Quantum Chemistry}, volume={115}, pages={647--655}, year={2015}, doi={10.1002/qua.24837}}

@article{Rohlfing2000, author={Rohlfing, Michael and Louie, Steven G.}, title={Electron-hole excitations and optical spectra from first principles}, journal={Physical Review B}, volume={62}, pages={4927--4944}, year={2000}, doi={10.1103/PhysRevB.62.4927}}

@article{RomanPerez2009, author={Rom{\'a}n-P{\'e}rez, Guillermo and Soler, Jos{\'e} M.}, title={Efficient Implementation of a van der Waals Density Functional: Application to Double-Wall Carbon Nanotubes}, journal={Physical Review Letters}, volume={103}, pages={096102}, year={2009}, doi={10.1103/PhysRevLett.103.096102}}

@article{Ruffieux2012, author={Ruffieux, Pascal and Cai, Jinming and Plumb, Nicholas C. and Patthey, Luc and Prezzi, Deborah and Ferretti, Andrea and Molinari, Elisa and Feng, Xinliang and Müllen, Klaus and Pignedoli, Carlo A. and Fasel, Roman}, title={Electronic Structure of Atomically Precise Graphene Nanoribbons}, journal={ACS Nano}, volume={6}, pages={6930--6935}, year={2012}, doi={10.1021/nn3021376}}

@article{Ruffieux2016, author={Ruffieux, Pascal and Wang, Shiyong and Yang, Bo and Sánchez-Sánchez, Carlos and Liu, Jia and Dienel, Thomas and Talirz, Leopold and Shinde, Prashant and Pignedoli, Carlo A. and Passerone, Daniele and Dumslaff, Tim and Feng, Xinliang and Müllen, Klaus and Fasel, Roman}, title={On-surface synthesis of graphene nanoribbons with zigzag edge topology}, journal={Nature}, volume={531}, pages={489--492}, year={2016}, doi={10.1038/nature17151}}

@article{Salpeter1951, author={Salpeter, E. E. and Bethe, H. A.}, title={A Relativistic Equation for Bound-State Problems}, journal={Physical Review}, volume={84}, pages={1232--1242}, year={1951}, doi={10.1103/PhysRev.84.1232}}

@article{Santos2021, author={dos Santos, Ramiro M. and Pereira Junior, Marcelo L. and Roncaratti, Luiz F. and Ribeiro Junior, Luiz A.}, title={Predicting the energetic stabilization of Janus-MoSSe/AlN heterostructures: A DFT study}, journal={Chemical Physics Letters}, volume={771}, pages={138465}, year={2021}, doi={10.1016/j.cplett.2021.138465}}

@article{Santos2025, author={dos Santos, Emanuel J. A. and Pereira, Marcelo L. and Tromer, Raphael M. and Galvão, Douglas S. and Ribeiro, Luiz A.}, title={Exploring the electronic and mechanical properties of the recently synthesized nitrogen-doped amorphous monolayer carbon}, journal={Nanoscale}, volume={17}, pages={7253--7263}, year={2025}, doi={10.1039/d4nr04305g}}

@article{Schelling2002, author={Schelling, Patrick K. and Phillpot, Simon R. and Keblinski, Pawel}, title={Comparison of atomic-level simulation methods for computing thermal conductivity}, journal={Physical Review B}, volume={65}, pages={144306}, year={2002}, doi={10.1103/PhysRevB.65.144306}}

@article{Schiros2012, author={Schiros, Theanne and Nordlund, Dennis and Pálová, Lucia and Prezzi, Deborah and Zhao, Liuyan and Kim, Keun Soo and Wurstbauer, Ulrich and Gutiérrez, Christopher and Delongchamp, Dean and Jaye, Cherno and Fischer, Daniel and Ogasawara, Hirohito and Pettersson, Lars G. M. and Reichman, David R. and Kim, Philip and Hybertsen, Mark S. and Pasupathy, Abhay N.}, title={Connecting Dopant Bond Type with Electronic Structure in N-Doped Graphene}, journal={Nano Letters}, volume={12}, pages={4025--4031}, year={2012}, doi={10.1021/nl301409h}}

@article{Shang2011, author={Shang, Honghui and Li, Zhenyu and Yang, Jinlong}, title={Implementation of screened hybrid density functional for periodic systems with numerical atomic orbitals: Basis function fitting and integral screening}, journal={The Journal of Chemical Physics}, volume={135}, pages={034110}, year={2011}, doi={10.1063/1.3610379}}

@article{Soler2002, author={Soler, Jos{\'e} M. and Artacho, Emilio and Gale, Julian D. and Garc{\'i}a, Alberto and Junquera, Javier and Ordej{\'o}n, Pablo and S{\'a}nchez-Portal, Daniel}, title={The SIESTA method for ab initio order-N materials simulation}, journal={Journal of Physics: Condensed Matter}, volume={14}, pages={2745--2779}, year={2002}, doi={10.1088/0953-8984/14/11/302}}

@article{Son2006, author={Son, Young-Woo and Cohen, Marvin L. and Louie, Steven G.}, title={Energy Gaps in Graphene Nanoribbons}, journal={Physical Review Letters}, volume={97}, pages={216803}, year={2006}, doi={10.1103/PhysRevLett.97.216803}}

@article{Talirz2016, author={Talirz, Leopold and Ruffieux, Pascal and Fasel, Roman}, title={On-Surface Synthesis of Atomically Precise Graphene Nanoribbons}, journal={Advanced Materials}, volume={28}, pages={6222--6231}, year={2016}, doi={10.1002/adma.201505738}}

@article{Talirz2017, author={Talirz, Leopold and Söde, Hajo and Dumslaff, Tim and Wang, Shiyong and Sanchez-Valencia, Juan Ramon and Liu, Jia and Shinde, Prashant and Pignedoli, Carlo A. and Liang, Liangbo and Meunier, Vincent and Plumb, Nicholas C. and Shi, Ming and Feng, Xinliang and Narita, Akimitsu and Müllen, Klaus and Fasel, Roman and Ruffieux, Pascal}, title={On-Surface Synthesis and Characterization of 9-Atom Wide Armchair Graphene Nanoribbons}, journal={ACS Nano}, volume={11}, pages={1380--1388}, year={2017}, doi={10.1021/acsnano.6b06405}}

@article{Thompson2022, author={Thompson, Aidan P. and Aktulga, H. Metin and Berger, Richard and Bolintineanu, Dan S. and Brown, W. Michael and Crozier, Paul S. and in 't Veld, Pieter J. and Kohlmeyer, Axel and Moore, Stan G. and Nguyen, Trung Dac and Shan, Ray and Stevens, Mark J. and Tranchida, Julien and Trott, Christian and Plimpton, Steven J.}, title={LAMMPS -- a flexible simulation tool for particle-based materials modeling at the atomic, meso, and continuum scales}, journal={Computer Physics Communications}, volume={271}, pages={108171}, year={2022}, doi={10.1016/j.cpc.2021.108171}}

@article{Tison2015, author={Tison, Yann and Lagoute, J{\'e}r{\^o}me and Repain, Vincent and Chacon, Cyril and Girard, Yann and Rousset, Sylvie and Joucken, Fr{\'e}d{\'e}ric and Sharma, Dimpy and Henrard, Luc and Amara, Hakim and Ghedjatti, Ahmed and Ducastelle, Fran{\c c}ois}, title={Electronic Interaction between Nitrogen Atoms in Doped Graphene}, journal={ACS Nano}, volume={9}, pages={670--678}, year={2015}, doi={10.1021/nn506074u}}

@article{Togo2015, author={Togo, Atsushi and Tanaka, Isao}, title={First principles phonon calculations in materials science}, journal={Scripta Materialia}, volume={108}, pages={1--5}, year={2015}, doi={10.1016/j.scriptamat.2015.07.021}}

@article{Togo2023, author={Togo, Atsushi}, title={First-principles Phonon Calculations with Phonopy and Phono3py}, journal={Journal of the Physical Society of Japan}, volume={92}, pages={012001}, year={2023}, doi={10.7566/JPSJ.92.012001}}

@article{Tries2020, author={Tries, Alexander and Osella, Silvio and Zhang, Pengfei and Xu, Fugui and Ramanan, Charusheela and Kläui, Mathias and Mai, Yiyong and Beljonne, David and Wang, Hai I.}, title={Experimental Observation of Strong Exciton Effects in Graphene Nanoribbons}, journal={Nano Letters}, volume={20}, pages={2993--3002}, year={2020}, doi={10.1021/acs.nanolett.9b04816}}

@article{Troullier1991, author={Troullier, N. and Martins, Jos{\'e} Lu{\'i}s}, title={Efficient pseudopotentials for plane-wave calculations}, journal={Physical Review B}, volume={43}, pages={1993--2006}, year={1991}, doi={10.1103/PhysRevB.43.1993}}

@article{vanDuin2001, author={van Duin, Adri C. T. and Dasgupta, Siddharth and Lorant, Francois and Goddard, William A.}, title={ReaxFF: A Reactive Force Field for Hydrocarbons}, journal={The Journal of Physical Chemistry A}, volume={105}, pages={9396--9409}, year={2001}, doi={10.1021/jp004368u}}

@article{Vanin2010, author={Vanin, M. and Mortensen, J. J. and Kelkkanen, A. K. and Garcia-Lastra, J. M. and Thygesen, K. S. and Jacobsen, K. W.}, title={Graphene on metals: A van der Waals density functional study}, journal={Physical Review B}, volume={81}, pages={081408}, year={2010}, doi={10.1103/PhysRevB.81.081408}}

@article{Wang2005, author={Wang, Feng and Dukovic, Gordana and Brus, Louis E. and Heinz, Tony F.}, title={The Optical Resonances in Carbon Nanotubes Arise from Excitons}, journal={Science}, volume={308}, pages={838--841}, year={2005}, doi={10.1126/science.1110265}}

@article{Wang2012Ngraphene, author={Wang, Haibo and Maiyalagan, Thandavarayan and Wang, Xin}, title={Review on Recent Progress in Nitrogen-Doped Graphene: Synthesis, Characterization, and Its Potential Applications}, journal={ACS Catalysis}, volume={2}, pages={781--794}, year={2012}, doi={10.1021/cs200652y}}

@article{Wang2017pyrazinePAH, author={Wang, Xiao-Ye and Richter, Marcus and He, Yuanqin and Bj{\"o}rk, Jonas and Riss, Alexander and Rajesh, Raju and Garnica, Manuela and Hennersdorf, Felix and Weigand, Jan J. and Narita, Akimitsu and Berger, Reinhard and Feng, Xinliang and Auw{\"a}rter, Willi and Barth, Johannes V. and Palma, Carlos-Andres and M{\"u}llen, Klaus}, title={Exploration of pyrazine-embedded antiaromatic polycyclic hydrocarbons generated by solution and on-surface azomethine ylide homocoupling}, journal={Nature Communications}, volume={8}, pages={1948}, year={2017}, doi={10.1038/s41467-017-01934-1}}

@article{Wang2018chiralGNR, author={Wang, Xiao-Ye and Urgel, Jos{\'e} I. and Barin, Gabriela Borin and Eimre, Kristjan and Di Giovannantonio, Marco and Milani, Alberto and Tommasini, Matteo and Pignedoli, Carlo A. and Ruffieux, Pascal and Feng, Xinliang and Fasel, Roman and M{\"u}llen, Klaus and Narita, Akimitsu}, title={Bottom-Up Synthesis of Heteroatom-Doped Chiral Graphene Nanoribbons}, journal={Journal of the American Chemical Society}, volume={140}, pages={9104--9107}, year={2018}, doi={10.1021/jacs.8b06210}}

@article{Wang2021quantum, author={Wang, Haomin and Wang, Hui Shan and Ma, Chuanxu and Chen, Lingxiu and Jiang, Chengxin and Chen, Chen and Xie, Xiaoming and Li, An-Ping and Wang, Xinran}, title={Graphene nanoribbons for quantum electronics}, journal={Nature Reviews Physics}, volume={3}, pages={791--802}, year={2021}, doi={10.1038/s42254-021-00370-x}}

@article{Wen2023, author={Wen, Ethan Chi Ho and Jacobse, Peter H. and Jiang, Jingwei and Wang, Ziyi and Louie, Steven G. and Crommie, Michael F. and Fischer, Felix R.}, title={Fermi-Level Engineering of Nitrogen Core-Doped Armchair Graphene Nanoribbons}, journal={Journal of the American Chemical Society}, volume={145}, pages={19338--19346}, year={2023}, doi={10.1021/jacs.3c05755}}

@article{Xu2014, author={Xu, Xiangfan and Pereira, Luiz F. C. and Wang, Yu and Wu, Jing and Zhang, Kaiwen and Zhao, Xiangming and Bae, Sukang and Tinh Bui, Cong and Xie, Rongguo and Thong, John T. L. and Hong, Byung Hee and Loh, Kian Ping and Donadio, Davide and Li, Baowen and Özyilmaz, Barbaros}, title={Length-dependent thermal conductivity in suspended single-layer graphene}, journal={Nature Communications}, volume={5}, pages={3689}, year={2014}, doi={10.1038/ncomms4689}}

@article{Yang2007Exciton, author={Yang, Li and Cohen, Marvin L. and Louie, Steven G.}, title={Excitonic Effects in the Optical Spectra of Graphene Nanoribbons}, journal={Nano Letters}, volume={7}, pages={3112--3115}, year={2007}, doi={10.1021/nl0716404}}

@article{Yang2007QP, author={Yang, Li and Park, Cheol-Hwan and Son, Young-Woo and Cohen, Marvin L. and Louie, Steven G.}, title={Quasiparticle Energies and Band Gaps in Graphene Nanoribbons}, journal={Physical Review Letters}, volume={99}, pages={186801}, year={2007}, doi={10.1103/PhysRevLett.99.186801}}

@article{Yazyev2007, author={Yazyev, Oleg V. and Helm, Lothar}, title={Defect-induced magnetism in graphene}, journal={Physical Review B}, volume={75}, pages={125408}, year={2007}, doi={10.1103/PhysRevB.75.125408}}

@article{Yazyev2010, author={Yazyev, Oleg V.}, title={Emergence of magnetism in graphene materials and nanostructures}, journal={Reports on Progress in Physics}, volume={73}, pages={056501}, year={2010}, doi={10.1088/0034-4885/73/5/056501}}

@article{Yousefi2020, author={Yousefi, Farrokh and Khoeini, Farhad and Rajabpour, Ali}, title={Thermal conductivity and thermal rectification of nanoporous graphene: A molecular dynamics simulation}, journal={International Journal of Heat and Mass Transfer}, volume={146}, pages={118884}, year={2020}, doi={10.1016/j.ijheatmasstransfer.2019.118884}}

@article{Yutomo2021, author={Yutomo, Erik Bhekti and Noor, Fatimah Arofiati and Winata, Toto}, title={Effect of the number of nitrogen dopants on the electronic and magnetic properties of graphitic and pyridinic N-doped graphene: a density-functional study}, journal={RSC Advances}, volume={11}, pages={18371--18380}, year={2021}, doi={10.1039/d1ra01095f}}

@article{Zhang2022, author={Zhang, Yong and Lu, Jianchen and Li, Yang and Li, Baijin and Ruan, Zilin and Zhang, Hui and Hao, Zhenliang and Sun, Shijie and Xiong, Wei and Gao, Lei and Chen, Long and Cai, Jinming}, title={On-Surface Synthesis of a Nitrogen-Doped Graphene Nanoribbon with Multiple Substitutional Sites}, journal={Angewandte Chemie International Edition}, volume={61}, pages={e202204736}, year={2022}, doi={10.1002/anie.202204736}}

@article{Zhao2011, author={Zhao, Liuyan and He, Rui and Rim, Kwang Taeg and Schiros, Theanne and Kim, Keun Soo and Zhou, Hui and Gutiérrez, Christopher and Chockalingam, S. P. and Arguello, Carlos J. and Pálová, Lucia and Nordlund, Dennis and Hybertsen, Mark S. and Reichman, David R. and Heinz, Tony F. and Kim, Philip and Pinczuk, Aron and Flynn, George W. and Pasupathy, Abhay N.}, title={Visualizing Individual Nitrogen Dopants in Monolayer Graphene}, journal={Science}, volume={333}, pages={999--1003}, year={2011}, doi={10.1126/science.1208759}}

@article{Son2007erratum, author={Son, Young-Woo and Cohen, Marvin L. and Louie, Steven G.}, title={Erratum: Energy Gaps in Graphene Nanoribbons [Phys. Rev. Lett. 97, 216803 (2006)]}, journal={Physical Review Letters}, volume={98}, pages={089901}, year={2007}, doi={10.1103/PhysRevLett.98.089901}}

@article{Dion2005erratum, author={Dion, M. and Rydberg, H. and Schr{\"o}der, E. and Langreth, D. C. and Lundqvist, B. I.}, title={Erratum: Van der Waals Density Functional for General Geometries [Phys. Rev. Lett. 92, 246401 (2004)]}, journal={Physical Review Letters}, volume={95}, pages={109902}, year={2005}, doi={10.1103/PhysRevLett.95.109902}}

@misc{sisl, author={Papior, Nick}, title={sisl: a toolbox for electronic structure calculations}, year={2024}, doi={10.5281/zenodo.597181}, note={Version 0.16.4}}

@article{Aparicio2025acsaem, author={Aparicio-Huacarpuma, Bill D. and Pereira Júnior, Marcelo Lopes and Silva, Alysson M. A. and Dias, Alexandre C. and Ribeiro Júnior, Luiz Antônio}, title={Solar Harvesting Efficiency of Janus M$_2$CTT$'$ (M = Y, Sc; T/T$'$ = Br, Cl, F) MXene Monolayers for Photovoltaic Applications}, journal={ACS Applied Energy Materials}, volume={8}, pages={6634--6644}, year={2025}, doi={10.1021/acsaem.5c00657}}

@article{Aparicio2025nanoscale, author={Aparicio-Huacarpuma, Bill D. and Pereira, Marcelo L. and Piotrowski, Mauricio J. and Rêgo, Celso R. C. and Guedes-Sobrinho, Diego and Ribeiro, Luiz A. and Dias, Alexandre C.}, title={Enhanced solar harvesting efficiency in nanostructured MXene monolayers based on scandium and yttrium}, journal={Nanoscale}, volume={17}, pages={13298--13310}, year={2025}, doi={10.1039/D4NR05505E}}

@article{fthenakis2022evaluating,
  title={Evaluating the performance of ReaxFF potentials for sp2 carbon systems (graphene, carbon nanotubes, fullerenes) and a new ReaxFF potential},
  author={Fthenakis, Zacharias G and Petsalakis, Ioannis D and Tozzini, Valentina and Lathiotakis, Nektarios N},
  journal={Frontiers in Chemistry},
  volume={10},
  pages={951261},
  year={2022},
  doi={10.3389/fchem.2022.951261},
  publisher={Frontiers Media SA}
}

@article{liao2011thermally,
  title={Thermally limited current carrying ability of graphene nanoribbons},
  author={Liao, Albert D and Wu, Justin Z and Wang, Xinran and Tahy, Kristof and Jena, Debdeep and Dai, Hongjie and Pop, Eric},
  journal={Physical Review Letters},
  volume={106},
  number={25},
  pages={256801},
  year={2011},
    doi={10.1103/PhysRevLett.106.256801},
  publisher={APS}
}

@article{ng2012molecular,
  title={A molecular dynamics study of the thermal conductivity of graphene nanoribbons containing dispersed Stone--Thrower--Wales defects},
  author={Ng, Teng Yong and Yeo, Jing Jie and Liu, ZS},
  journal={Carbon},
  volume={50},
  number={13},
  pages={4887--4893},
  year={2012},
  doi={10.1016/j.carbon.2012.06.017},
  publisher={Elsevier}
}

\end{document}